\documentclass[fleqn,usenatbib]{mnras}

\usepackage{newtxtext,newtxmath}
\usepackage{subfig}

\usepackage{ae,aecompl}

\usepackage{graphicx}	
\usepackage{amsmath}	
\usepackage{amssymb}	

\title[IllustrisTNG: Build-up of spheroids and discs]{Morphology and star formation in IllustrisTNG: \\the build-up of spheroids and discs}

\author[Tacchella et al.]{Sandro Tacchella$^{1}$\thanks{E-mail: sandro.tacchella@cfa.harvard.edu},
Benedikt Diemer$^{1}$\thanks{NHFP Einstein Fellow},
Lars Hernquist$^{1}$,
Shy Genel$^{2,3}$,
\newauthor
Federico Marinacci$^{4}$,
Dylan Nelson$^{5}$,
Annalisa Pillepich$^{6}$,
\newauthor
Vicente Rodriguez-Gomez$^{7}$,
Laura V. Sales$^{8}$,
Volker Springel$^{5}$, and
\newauthor
Mark Vogelsberger$^{9}$
\\
\\
$^{1}$Center for Astrophysics $|$ Harvard \& Smithsonian, 60 Garden St, Cambridge, MA 02138, USA\\
$^{2}$Center for Computational Astrophysics, Flatiron Institute, 162 Fifth Avenue, New York, NY 10010, USA\\
$^{3}$Columbia Astrophysics Laboratory, Columbia University, 550 West 120th Street, New York, NY 10027, USA\\
$^{4}$Department of Physics \& Astronomy, University of Bologna, via Gobetti 93/2, 40129 Bologna, Italy\\
$^{5}$Max-Planck-Institut f\"{u}r Astrophysik, Karl-Schwarzschild-Strasse 1, D-85741 Garching bei M\"{u}nchen, Germany\\
$^{6}$Max-Planck-Institut f\"{u}r Astronomie, K\"{o}nigstuhl 17, 69117 Heidelberg, Germany\\
$^{7}$Instituto de Radioastronom\'ia y Astrof\'isica, Universidad Nacional Aut\'onoma de M\'exico, Apdo. Postal 72-3, 58089 Morelia, Mexico\\
$^{8}$Department of Physics and Astronomy, University of California, Riverside, 900 University Avenue, Riverside, CA 92521, USA\\
$^{9}$Kavli Institute for Astrophysics and Space Research, Massachusetts Institute of Technology, Cambridge, MA 02139, USA\\
}

\date{\textit{Draft version: \today}}
\pubyear{2019}
\begin{document}
\label{firstpage}
\pagerange{\pageref{firstpage}--\pageref{lastpage}}
\maketitle

\begin{abstract}
Using the IllustrisTNG simulations, we investigate the connection between galaxy morphology and star formation in central galaxies with stellar masses in the range $10^9-10^{11.5}~\mathrm{M}_{\odot}$. We quantify galaxy morphology by a kinematical decomposition of the stellar component into a spheroidal and a disc component (spheroid-to-total ratio, S/T) and by the concentration of the stellar mass density profile ($C_{82}$). S/T is correlated with stellar mass and star-formation activity, while $C_{82}$ correlates only with stellar mass. Overall, we find good agreement with observational estimates for both S/T and $C_{82}$. Low- and high-mass galaxies are dominated by random stellar motion, while only intermediate-mass galaxies ($M_{\star}\approx10^{10}-10^{10.5}~\mathrm{M}_{\odot}$) are dominated by ordered rotation. Whereas higher mass galaxies are typical spheroids with high concentrations, lower-mass galaxies have low concentration, pointing to different formation channels. Although we find a correlation between S/T and star-formation activity, in the TNG model galaxies do not necessarily change their morphology when they transition through the green valley or when they cease their star formation, this depending on galaxy stellar mass and morphological estimator. Instead, the morphology (S/T and $C_{82}$) is generally set during the star-forming phase of galaxies. The apparent correlation between S/T and star formation arises because earlier forming galaxies had, on average, a higher S/T at a given stellar mass. Furthermore, we show that mergers drive in-situ bulge formation in intermediate-mass galaxies and are responsible for the recent spheroidal mass assembly in the massive galaxies with $M_{\star}>10^{11}~\mathrm{M}_{\odot}$. In particular, these massive galaxies assemble about half of the spheroidal mass while star-forming and the other half through mergers while quiescent.
\end{abstract}

\begin{keywords}
galaxies: evolution -- galaxies: formation -- galaxies: fundamental parameters -- galaxies: high-redshift -- galaxies: structure
\end{keywords}



\section{Introduction}
\label{sec:intro}

Today's galaxy population exhibits a large structural diversity that depends on stellar mass ($M_{\star}$), star-formation activity, and environment \citep{strateva01,baldry06,schawinski14,cappellari16}. Although tremendous progress has been made by measuring galaxy morphology out to redshift of $z\approx3$, it remains elusive how the spheroidal and disc components form and evolve within galaxies. Specifically, the relative contributions of mergers, secular evolution, violent disc instabilities, accretion of gas with low angular momentum, and other mechanisms to the observed structural diversity are unclear. 

The \textit{Hubble Space Telescope} enables us to study the rest-frame optical morphologies of high-$z$ galaxies and has revealed that star-forming galaxies at $z\sim2$ often appear more regular at longer than at shorter wavelengths \citep[e.g.,][]{toft07, elmegreen09, forster-schreiber11b, wuyts12}. Their kinematics are mostly constrained by warm ionized gas as traced by the H$\alpha$ emission line, while the stellar kinematics remain elusive. These H$\alpha$-based estimates point to a high fraction ($>50\%$) of discs, characterized by high intrinsic local velocity dispersions of $\sigma_0\approx25-100~\mathrm{km}~\mathrm{s}^{-1}$ \citep[e.g.,][]{forster-schreiber06b, genzel08, law09, wisnioski15, wisnioski18, stott16, forster-schreiber18_SINS, ubler18}. Combined with the relatively small scatter in the star-formation rates (SFRs) at fixed stellar mass (the star-forming main sequence; \citealt{noeske07b,elbaz07,whitaker12}), the regular morphology and disc-like kinematics suggest that galaxies grow in a quasi-steady state of gas inflow, outflow, and consumption, while mergers play a subdominant role \citep{bouche10,hopkins10a, genzel10, lilly13_bathtube, forbes14a, nelson16_insideout, tacchella16_MS}. 

Nevertheless, there is plenty of evidence that normal, $z\sim2$ star-forming galaxies on the main sequence have built and are building their bulges: the bulge-to-total ratio smoothly increases with stellar mass in both star-forming and quiescent galaxies \citep{bruce14_BD, lang14, bluck14, mendel15, dimauro19}, some star-forming galaxies show signs of a nuclear starburst \citep{barro13, barro14, nelson14_nature, tadaki17}, and massive star-forming galaxies have a central stellar mass density comparable to local quiescent systems \citep{van-dokkum14_dense_cores, tacchella15_sci, barro17}. Furthermore, there is growing evidence that quiescent galaxies at $z=1-3$ are disc-like and rotating \citep[e.g.,][]{chang13, van-de-sande13, newman15, toft17}. Together with the mature bulges present in massive star-forming galaxies, this indicates that probably no significant bulge growth and change in the kinematics needs to take place when galaxies halt their star formation, i.e. that ``discs'' do \textit{not} need to be transformed into spheroids \citep{tacchella15_sci, tacchella18_dust}. Contrarily, observational studies using abundance matching to connect galaxy populations through cosmic time find that the growth of galaxy bulges in massive galaxies corresponds to a rapid decline in the galaxy gas fractions and SFRs \citep[e.g.,][]{van-dokkum10, patel13, papovich15}.

In the local Universe, observational constraints are much more detailed, including the shape, dynamics, and stellar population of complete samples of galaxies \citep[see review by][]{cappellari16}. The quiescent galaxies can be classified into two main classes using stellar kinematics: slow and fast rotators \citep[e.g.,][]{davis83, kormendy96, emsellem07, veale17, smethurst18}. The slow rotators are rare and dominate only at $M_{\star}>2\times10^{11}~\mathrm{M}_{\odot}$ \citep{illingworth77, binney78}. These galaxies assemble near the centre of massive haloes (e.g., clusters and massive groups) and bear clear signatures of a gas-poor formation processes \citep[gas-poor mergers;][]{naab03, cox06, bezanson09, penoyre17, faisst17_size, schulze18, zahid19}. The more typical population of quiescent galaxies, at a mass scale of $M_{\star}\sim10^{11}~\mathrm{M}_{\odot}$, consists of fast rotators with disc-like isophotes \citep{bender88}, steep nuclear light profiles \citep{faber97}, and steep metallicity gradients \citep{carollo93}: all features that indicate a gas-rich formation process, as seen in the star-forming population at $z=1-3$. 

Traditionally, stellar spheroids are thought to arise from dissipationless accumulation of previously formed stars in mergers \citep[e.g.,][]{cole00}, while discs are thought to arise from star formation in the dissipational collapse of high angular momentum gas \citep[e.g.,][]{fall80}. In this picture, the contributions from mass formed in spheroids and in discs could tell us about the balance between the different modes of galaxy formation. However, high-$z$ observations and numerical simulations show that the spheroidal component in the centres of galaxies may initially form at high redshifts. This high-$z$ bulge formation is driven by a variety of processes that conspire to rapidly fuel gas into the central region \citep[summarily called ``gas compaction'';][]{dekel14_nugget, zolotov15,tacchella16_profile}. These processes include gas inflow enabled by disc instabilities, counter-rotating streams, clump migration, and mergers \citep[e.g.][]{hernquist89, noguchi99, dekel09b, bournaud11, sales12, wellons15, wellons16}. Even when galaxies cease their star formation, their spheroidal component can still be augmented by gas-poor mergers and perturbations such as tidal interactions from neighbouring galaxies \citep{naab09, bekki11, oser12}. Finally, the time and type of merger also impacts the galaxies' structure \citep[e.g.,][]{barnes92b, hernquist92, hernquist93, naab03, cox06, rodriguez-gomez17}. Specifically, mergers can spin galaxies up or down \citep{naab14,lagos17,penoyre17}, and can lead to the formation of a diffuse halo \citep{brook11, pillepich15}. 

Given this rich variety of pathways by which spheroids form, the evolution of galaxy structure provides invaluable constraints on key physical processes (that are uncertain and thus often captured in ``sub-grid models'', e.g., for black hole feedback, stellar feedback, and the ISM). For example, past numerical simulations have often struggled to prevent excessive spheroid formation \citep[e.g.,][]{navarro94, navarro00, van-den-bosch01, donghia06, scannapieco12}. Stellar feedback is required to remove low angular momentum gas, also during merger-induced starbursts \citep[e.g.,][]{governato09, governato10, brook11, brook12, christensen14, zjupa17}. Similarly, black hole feedback might be necessary to further suppress spheroid formation and to prevent disc re-growth at later times \citep[e.g.,][]{genel15, dubois16, sparre17}.

To understand the physics driving the evolution of morphology, a number of studies have used cosmological hydrodynamical simulations \citep[e.g.,][]{dubois16, rodriguez-gomez17, correa17, martin18, wright19, trayford19, thob19}. For example, using the EAGLE simulations \citep{schaye15,crain15}, \citet{correa19} note that the time when galaxies move onto the red sequence depends on their morphology, with only a weak connection between transformations in colour and morphology. Similarly, \citet{clauwens18} put forward that the morphology of EAGLE galaxies is primarily set by their stellar mass. They distinguish three phases of galaxy formation, where the low-mass phase is dominated by disorganized growth, the intermediate-mass phase corresponds to disc growth, and the high-mass phase transforms discs into spheroids. This final transformation is driven more by the build-up of spheroids than by the destruction of discs. 

In this paper, we investigate the formation of spheroids and discs in the IllustrisTNG simulations \citep{pillepich18_cluster, nelson18_color,springel18,naiman18,marinacci18}. Previously, \citet{pillepich19} investigated the structural and kinematical evolution of star-forming galaxies across cosmic time ($0.5<z<6$) in the TNG50 simulation. They find that the vast majority of star-forming galaxies are rotationally supported gaseous discs for most cosmic epochs ($M_{\star}> 10^9~\mathrm{M}_{\odot}$, $z\la4-5$), while the underlying stellar component is much less rotationally supported. We build on their analysis, but focus on how and when the morphology of galaxies is determined, and how this process relates to star-formation activity. We include quiescent galaxies to $z=0$ and thus focus on the larger TNG100 simulation. Specifically, we address the following questions. When and how do spheroidal components form? When are galaxies able to efficiently form a disc? Is there any morphological transformation when galaxies cease their star formation?

The paper is structured as follows. We give an overview of the IllustrisTNG simulations, the galaxy sample, and our morphological indicators in Section~\ref{sec:illustrisTNG}. In Section~\ref{sec:spheroids_in_pop}, we discuss how morphology is correlated with stellar mass and star formation and show how and when spheroids and discs form in IllustrisTNG. Section~\ref{sec:consequences} highlights a few consequences for the $z=0$ galaxy population, in particular concerning the ex-situ stellar mass distribution in the galaxy population and within galaxies. Section~\ref{sec:discussion} discusses the emerging picture and highlights the caveats of this analysis. Finally, we summarize and conclude in Section~\ref{sec:conclusion}. Additionally, we present an extended appendix in which we discuss the numerical resolution of the simulations (Appendix~\ref{app:resolution}), compare different morphological tracers (Appendix~\ref{app:diff_morph}), consider the morphology of galaxies on the star-forming main sequence (Appendix~\ref{app:MS_morph}), and show results for the original Illustris simulation (Appendix~\ref{app:illustris}).

\section{Simulation Data}
\label{sec:illustrisTNG}

\begin{table}
	\centering
	\caption{The most important properties of the three simulations analysed in this work. TNG100(-1) is the fiducial simulation, while TNG100-2 and TNG100-3 are lower resolution incarnations of the same box. The given parameters are: the simulated volume and box side-length (both in co-moving units); the number of initial gas cells and dark matter particles; the mean baryon and dark matter particle mass resolution in solar masses; the minimum allowed adaptive gravitational softening length for gas cells (co-moving Plummer equivalent); and the redshift zero softening of the dark matter and stellar components in physical units.}
	\label{tab:TNG100}
	\begin{tabular}{lclll} 
		\hline \hline
 Run & & TNG100(-1) & TNG100-2 & TNG100-3 \\ \hline
 Volume & [\,Mpc$^3$\,] & $110.7^3$ & $110.7^3$ & $110.7^3$ \\
 $L_{\rm box}$ & [\,Mpc/$h$\,] & 75 & 75 & 75 \\
 $N_{\rm GAS}$ & - & $1820^3$ & $910^3$ & $455^3$ \\
 $N_{\rm DM}$ & - & $1820^3$ & $910^3$ & $455^3$ \\
 $m_{\rm baryon}$ & [\,M$_\odot$\,] & $1.4 \times 10^6$ & $1.1 \times 10^7$ & $8.9 \times 10^7$ \\
 $m_{\rm DM}$ & [\,M$_\odot$\,] & $7.5 \times 10^6$ & $5.6 \times 10^7$ & $4.8 \times 10^8$ \\
 $\epsilon_{\rm gas,min}$ & [\,pc\,] & 185 & 370 & 738 \\
 $\epsilon_{\rm DM,stars}^{z=0}$ & [\,kpc\,] & 0.74 & 1.48 & 2.95 \\
		\hline
	\end{tabular}
\end{table}

\subsection{Description of the simulations}

The IllustrisTNG (The Next Generation; \citealt{nelson18_color, marinacci18, naiman18, pillepich18_cluster, springel18})\footnote{ \url{http://www.tng-project.org}} simulations build upon the successes of the Illustris galaxy formation model \citep{vogelsberger14, vogelsberger14a, genel14, sijacki15, nelson15_illustris}, and include prescriptions for star formation, stellar evolution, chemical enrichment, primordial and metal-line cooling, stellar feedback, galactic outflows, and black-hole formation, growth, and multimode feedback \citep[see][for a complete description]{weinberger17,pillepich18}.

The TNG project is made up of three simulation volumes: TNG50, TNG100, and TNG300. Here, we focus on TNG100 as a compromise between volume and resolution. The details of TNG100 are given in Table~\ref{tab:TNG100}. The fiducial TNG100 simulation includes 2$\times$1820$^3$ resolution elements in a $\sim100$ Mpc (comoving) box. The baryon mass resolution is $1.4\times10^6~\mathrm{M}_{\odot}$, the gravitational softening length of the dark matter and stars is 0.7  kpc at $z=0$, and the gas component has an adaptive softening with a minimum of 185 comoving parsecs. In order to address resolution effects, we analyse two lower resolution runs, TNG100-2 and TNG100-3 (Table~\ref{tab:TNG100}), in Appendix~\ref{app:resolution}. 

The parameters of the TNG model have been chosen to roughly reproduce observations of the SFR density as a function of comic time, the galaxy stellar mass function at $z=0$, the present-day stellar-to-halo mass relation, galaxy stellar sizes. halo gas fractions, and the black-hole mass to galaxy or halo mass relation. The TNG model fixes two shortcomings of the original Illustris simulation that are critical for our purposes: the stellar sizes of galaxies were larger than observed by a factor of a few for $M_{\star}\la10^{10.7}~\mathrm{M}_{\odot}$ \citep{pillepich18, genel18}, and the colour distribution showed only a weak bimodality between red and blue galaxies \citep{nelson18_color}. These issues were addressed by modifications to galactic winds and to the growth and feedback of supermassive black holes. As a result, galactic winds in IllustrisTNG are generally faster and more effective at preventing star formation at all masses and times. We note, however, that the simulations were not tuned to reproduce the galaxy morphologies we consider in this paper. Thus, they not only represent a test of the TNG physics recipes, but also constitute predictions for and interpretations of observations.

\subsection{Galaxy sample}

Galaxies and their haloes are identified as gravitationally bound substructures using the \textsc{Subfind} algorithm \citep{springel01c} and are then linked over time with the \textsc{SubLink} merger tree algorithm \citep{rodriguez-gomez15}. We investigate only galaxies with a stellar mass ($M_{\star}$) of $M_{\star}>10^{9}~\mathrm{M}_{\odot}$ in order to ensure a resolution of at least $\approx 1000$ stellar particles. Furthermore, we focus on central galaxies in order to minimize the effects of environment (such as stripping), which are beyond the scope of this work (see Joshi et al. in preparation for an analysis of morphological transformations induced by dense environments). Finally, we exclude a few objects with abnormal ratios of stellar to dark matter mass. Specifically we exclude galaxies with less than 5\% dark matter mass, thereby removing less than 0.01 \% of the galaxies (see also \citealt{nelson19_dr}). These selections result in $M_{\star}$-selected samples of 11,579, 10,939, and 8,294 galaxies at $z=0$, $z=1$, and $z=2$, respectively. As we wish to compare galaxy properties to observations, we measure all physical properties within an aperture of 3 times the 3D stellar half-mass radius ($R_{\rm M}$).

\subsection{Definition of star-forming and quiescent galaxies}
\label{subsec:SF_definition}

We estimate the SFR from the instantaneous gas properties, which we regard as the true SFR of the galaxies. Again, we measure the SFR within $3 R_{\rm M}$. A larger aperture does not change our results since the SFR profiles are usually centrally peaked. The specific star formation rate (sSFR) is the ratio between SFR and $M_{\star}$. To classify galaxies as star-forming, transitional, and quiescent, we compare their sSFR to the Hubble time, $t_{\rm H}(z)$, at the redshift of interest. Specifically, we identify quiescent galaxies as having an sSFR lower than 
\begin{equation}
    \mathrm{sSFR}_{\rm quiescent}(z) = \frac{1}{6 \times t_{\rm H}(z)} \,.
    \label{eq:ssfr_quiescent}
\end{equation}
This criterion implies that the stellar mass of quiescent galaxies effectively does not grow due to star formation, i.e., that the stellar mass grows by less than an $e$-fold (roughly a factor 2) within 6 Hubble times. Furthermore, we define transition galaxies to have $\mathrm{sSFR}_{\rm quiescent} < \mathrm{sSFR} <  \mathrm{sSFR}_{\rm transition}$, where
\begin{equation}
    \mathrm{sSFR}_{\rm transition}(z) = \frac{1}{2 \times t_{\rm H}(z)}.
    \label{eq:ssfr_transition}
\end{equation}
Since the star-forming main sequence in IllustrisTNG is roughly linear \citep{donnari19} and the normalization roughly evolves proportional to the Hubble time, these cuts are very similar to cuts relative to the main sequence, as shown in Figure~\ref{fig:MS_morph}.

A commonly used alternative definition of quiescence is based on colour. We use the $(g-r)$ colour as an additional indicator for star-formation activity, though our main distinction between star-forming and quiescent galaxies is based on sSFR. We compute the $(g-r)$ colour from all gravitationally bound stellar particles that belong to a galaxy, ignoring the effects of dust. As expected, we find a tight correlation where more quiescent galaxies exhibit redder colours. The key difference between sSFR and colour is that the former is sensitive to shorter timescales. Switching to a colour-based definition of quiescence changes our results quantitatively, but not qualitatively. 

\subsection{Measurement of morphology}
\label{subsec:morph_definition}

\begin{figure*}
	\includegraphics[width=\textwidth]{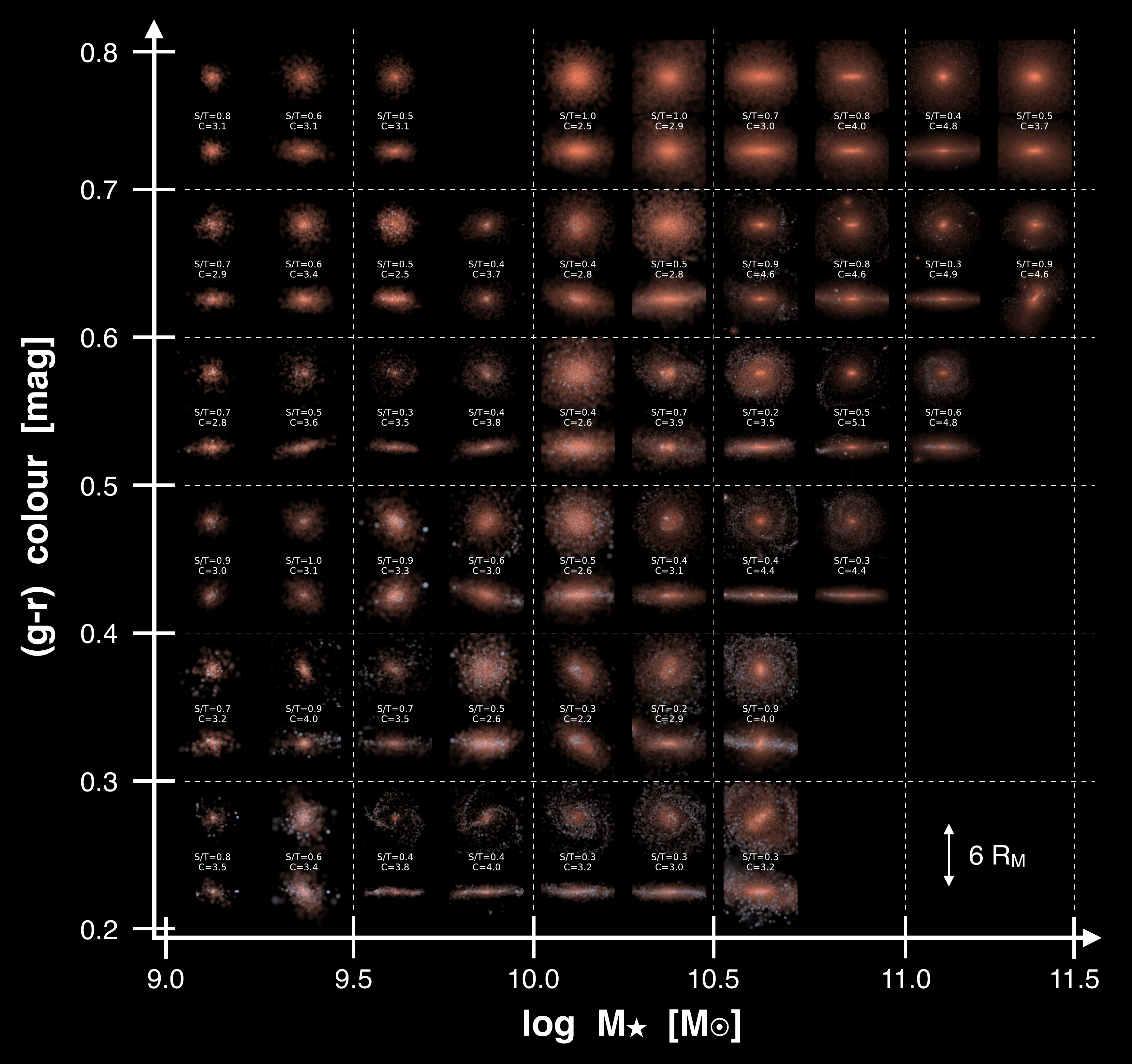}
    \caption{A visual demonstration of the diversity of galaxy morphology in the ($g-r$) colour$-M_{\star}$ plane. For each bin in ($g-r$) colour and $M_{\star}$, we randomly pick a galaxy and plot its face-on (above the label) and edge-on (below the label) projection. Each image is a colour-composite representing the stellar luminosity in the r-g-B bands, and is centred at the most bound particle in the galaxy. Each image has a total side length of 6 stellar half-mass (3D) radii. The label indicates the spheroid-to-total ratio (S/T) and concentration ($C_{82}$) of each galaxy. The main intention of this figure is to demonstrate the large diversity of visual morphology in this simulation. More massive and redder galaxies are more concentrated and have a more significant spheroidal component. We explore these trends further in Figure~\ref{fig:morphology_col_mass}.}
    \label{fig:stamps}
\end{figure*}

Observationally, the structure of galaxies, meaning their bulge-to-total ratio, size, and concentration, can be quantified in different ways. Most straightforwardly, the light distribution can be described either by non-parametric indicators or by a parametric model fit (e.g., S\'{e}rsic profile). Based on these descriptions, size, concentration, and bulge-to-total ratios can be estimated. Similarly, one can correct for the mass-to-light ratio variations and estimate these quantities directly from the stellar mass distribution \citep[e.g.,][]{wuyts12, lang14, cibinel15, tacchella15_sci, tacchella17_S1, mosleh17}. An alternative estimate of the structure of galaxies can be based on kinematics: ordered-to-random motion and the flattening of galaxies are two commonly used proxies for their angular momentum \citep[e.g.,][]{binney05,cappellari07}. A more sophisticated approach is to construct orbit-superposition \citet{schwarzschild79} models that simultaneously fit the observed surface brightness and stellar kinematics \citep[e.g.,][]{zhu18}. This makes it possible to determine the fraction of stellar mass in the kinematically cold (disc) and kinematically hot (spheroid) component. At least for massive galaxies at $z=0$, it has been shown that the photometric bulge-disc decompositions and the kinematical estimates from integral field spectroscopy yield remarkably consistent conclusions \citep{kormendy12,kormendy16}. 

In numerical simulations, estimating the structure of galaxies from kinematics is simpler than estimating it from the light distribution, since the latter involves forward modelling. Here, we focus on two different morphological indicators that are based on kinematics and the mass distribution, respectively. The first indicator is the spheroid-to-total ratio (S/T). We define the spheroidal component as the mass of all stellar particles with $j_z/j<0.7$ within 3 times $R_{\rm M}$. Here, $j$ is the total angular momentum of the particle and $j_z$ the component that is aligned with the galaxy's rotation (as defined by the total angular momentum of all stellar particles bound to the galaxy). This definition slightly underestimates the spheroidal component because a purely isotropic system would contain $\sim15\%$ of orbits with $j_z/j>0.7$. We correct for this effect by applying a global correction factor of $15\%$ to the spheroidal component. We have also considered definitions such as the fraction of kinetic energy in rotation $\kappa_{\rm rot}$ \citep{sales12}, twice the mass of all stellar particles that counter-rotate, and a different definition of the rotation of stellar particles. The different measures of S/T correlate very strongly with each other, indicating that the exact method for calculating S/T is not important for our purposes (Appendix~\ref{app:diff_morph} and Figure~\ref{fig:app_diff_sph}).

Our second morphological indicator is based on the 3D stellar mass density profile. In particular, we consider the concentration, defined as $C_{82}=5\times\log_{10}(r_{80}/r_{20})$, where $r_{80}$ and $r_{20}$ are the radii enclosing 80\% and 20\% of the stellar mass, respectively. This parameter is known to strongly correlate with the morphological type of galaxies \citep{kent85, bershady00, lotz04}. As before, we measure $r_{80}$ and $r_{20}$ within 3 times $R_{\rm M}$. Increasing the aperture leads to an overall increase of $C_{82}$ values, but the qualitative trends with $M_{\star}$, SFR, and colour remain unchanged. Similarly, estimating $C_{82}$ from the 2D instead of 3D stellar mass profile leads to no changes in our results. 

Our two indicators correlate weakly for low-mass galaxies, which exhibit a large diversity in S/T but $C_{82}\approx3$. Towards higher masses, $C_{82}$ and S/T correlate more tightly (Appendix~\ref{app:diff_morph} and Figure~\ref{fig:app_C_vs_ST}). We also present a detailed discussion concerning the effects of numerical resolution in Appendix~\ref{app:resolution} and discuss how $C_{82}$ and S/T are related to light-based measurements in Appendix~\ref{app:diff_morph}. 

Figure~\ref{fig:stamps} visually demonstrates the large diversity of galaxy morphology in the ($g-r$) colour$-M_{\star}$ plane. For each bin in ($g-r$) colour and $M_{\star}$, we randomly pick a galaxy and plot its face-on (above the label) and edge-on (below the label) projection. Each image shows a colour-composite representing the stellar luminosity in the r-g-B bands  \citep{genel18}. Massive and redder galaxies are clearly more concentrated and have a more significant spheroidal component. This visual impression is consistent with, and expands upon, the findings of \citet[][their figures 12 and 13]{nelson18_color}. We return to the colour$-M_{\star}-$morphology relation in Section~\ref{sec:spheroids_in_pop}.



\section{Spheroids and discs: formation and evolution}
\label{sec:spheroids_in_pop}

In this section, we consider the prevalence of spheroids and discs in the IllustrisTNG galaxy population. We first investigate how morphology depends on stellar mass and star formation, finding that TNG galaxies roughly follow the observed correlations. We then focus on the build-up of the spheroidal and disc components in galaxies over time, highlighting that both stellar mass \textit{and} cosmic time play key roles. 

\subsection{Dependence of morphology on stellar mass}
\label{subsec:mass_dependence}

\begin{figure}
	\includegraphics[width=\columnwidth]{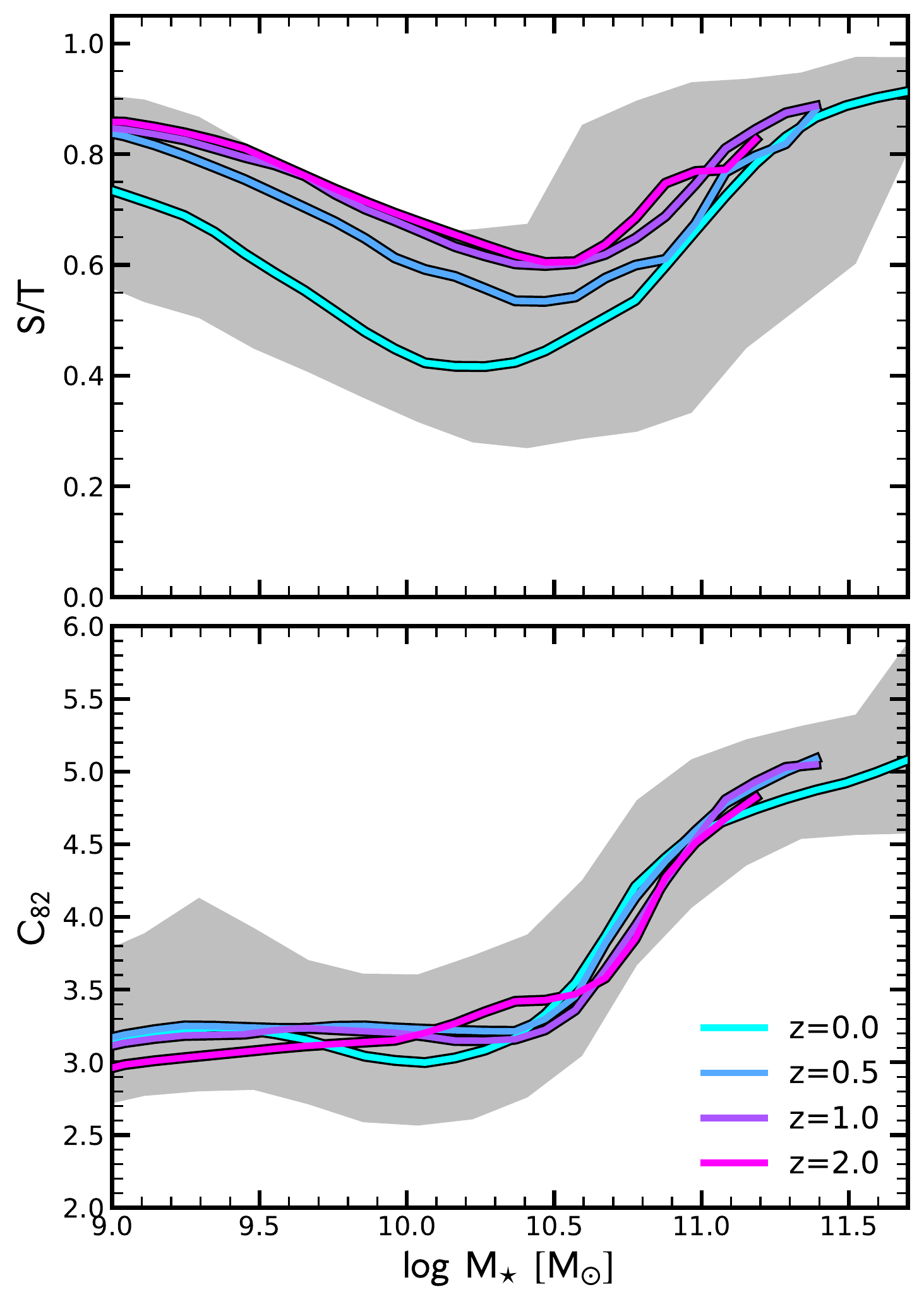}
    \caption{Spheroid-to-total ratio (S/T; top) and concentration of the stellar mass density profile ($C_{82}$; bottom) as a function of $M_{\star}$ and redshift. The cyan, blue, purple and pink lines show the median relations of S/T$-M_{\star}$ and $C_{82}-M_{\star}$ at $z=0.0, 0.5, 1.0,$ and 2.0, respectively. The shaded region shows the 1$\sigma$ scatter at $z=0.0$. Both morphological tracers agree in that massive galaxies are typical spheroids with a high concentration and hot stellar kinematics, while intermediate-mass galaxies are typical discs with a low concentration and a significant kinematically cold component. Conversely, at low stellar masses, S/T and $C_{82}$ diverge: those galaxies are kinematically hot although their concentration is low. Furthermore, while the $C_{82}-M_{\star}$ relation relation stays roughly constant with cosmic time, S/T decreases significantly from $z=0.5$ to $z=0.0$, implying efficient disc growth at late times.}
    \label{fig:morphology_vs_M}
\end{figure}

Figure~\ref{fig:morphology_vs_M} shows S/T and $C_{82}$ as a function of $M_{\star}$ and redshift. The trends of S/T and $C_{82}$ are similar at $M_{\star}>10^{10}~\mathrm{M}_{\odot}$. Typical disc galaxies with low S/T ($<0.5$) and low $C_{82}$ ($<3.5$) live in the mass range $M_{\star}\approx10^{10}-10^{11}~\mathrm{M}_{\odot}$ at $z=0$. More massive galaxies are typical spheroids with high S/T values of $>0.7$ and concentration measurements of $C_{82}\approx4.5-5.5$, which corresponds to a S\'{e}rsic index of $n\approx4$. Low-mass galaxies ($M_{\star}<10^{10}~\mathrm{M}_{\odot}$) show a significant difference between S/T and $C_{82}$: the kinematics of these galaxies is dominated by random motion, though their stellar mass distribution is well described by a low concentration, $C_{82}\approx3$, corresponding to a \citet{sersic68} index of $n\approx1$. We discuss these low-mass galaxies in detail in Section~\ref{subsec:lowM_galaxies}.

Figure~\ref{fig:morphology_vs_M} also demonstrates that there is little redshift evolution in the $C_{82}-M_{\star}$ relation. At all redshifts, high-mass galaxies have high concentration and low-mass galaxies have low concentration. The mass range where the concentration transitions is fixed at  $M_{\star}\approx10^{10.5}-10^{10.8}~\mathrm{M}_{\odot}$ at all times. Interestingly, this mass scale is comparable to the mass at which central galaxies in IllustrisTNG typically cease their star formation. Conversely, the S/T$-M_{\star}$ relation evolves significantly with redshift: low- and intermediate-mass galaxies have a decreasing S/T with cosmic time, in particular between $z=1$ and $z=0$. Moreover, the $M_{\star}$-location of the ``minimum'' in S/T evolves to lower $M_{\star}$ towards $z=0$. This trend indicates that, at a given stellar mass, galaxies build up their kinematically cold components over time. This is consistent with the idea of disc assembly as extensively explored in TNG50 in \citet{pillepich19} for both the stellar and gaseous structures of star-forming galaxies. This picture is qualitatively consistent with observations \citep[e.g.,][]{zhang19, van-der-wel14}, where the overall oblateness (or ``disciness'') of galaxies increases with cosmic time. 

Previous work by \citet{clauwens18} finds negligible redshift evolution of S/T$-M_{\star}$ relation in the EAGLE simulations. They define the mass of the spheroidal component as twice the mass of counter-rotating stars, which is the same as our S/T$_{\rm neg}$ mentioned above and in Appendix~\ref{app:diff_morph}. Using S/T$_{\rm neg}$ instead of our fiducial S/T measurement does not change our finding significantly: we still find a strong redshift evolution at $M_{\star}<10^{10.5}~\mathrm{M}_{\odot}$, while above this mass scale there is less evolution. This difference in the definition of S/T can therefore explain some of the difference between our work and the one by \citet{clauwens18}, but not all of it. It would be interesting to understand this in more detail in the future, analysing both simulations together in the same way.

\subsection{Comparison with observations}
\label{subsec:comparison_obs}

\begin{figure}
	\includegraphics[width=\columnwidth]{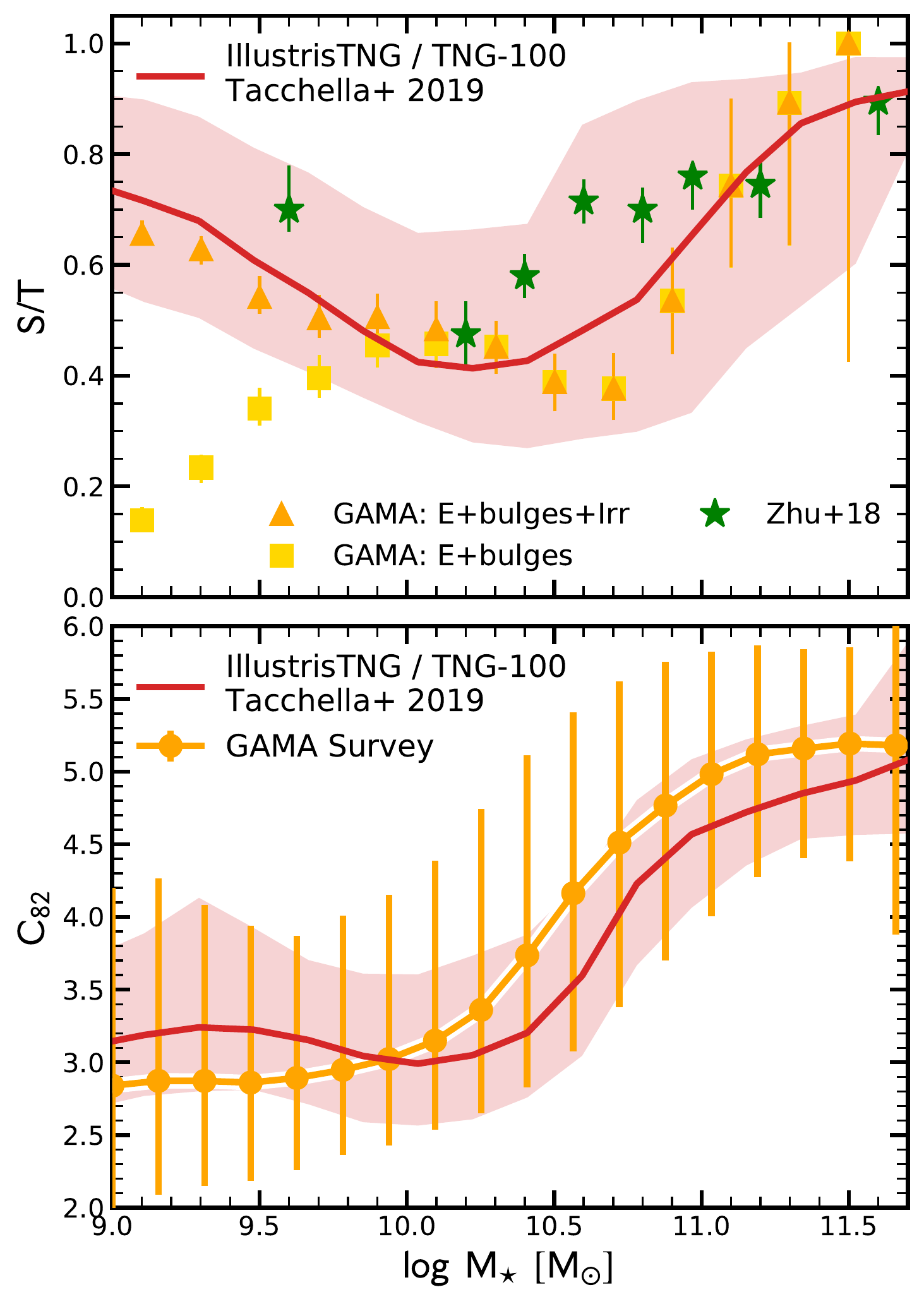}
    \caption{Comparison of S/T (top) and $C_{82}$ (bottom) to observations at $z=0$. The solid red lines in both panels show the running median of the IllustrisTNG galaxies, while the shaded regions indicates the $1\sigma$ scatter. The observational data are taken from \citet{zhu18} and the GAMA survey \citep{kelvin12, moffett16, lange16}. Top panel: the estimate by \citet{zhu18} stems from stellar orbit modelling and is comparable to our kinematics-based S/T estimate. The comparison to the light-based estimate from the GAMA survey should be interpreted with care, since the results at low masses depend significantly on whether one counts irregular galaxies as spheroids (triangles) or not (squares). Bottom panel: the concentration estimates from the GAMA survey stem from the $K$-band (errorbars indicate $1\sigma$ scatter) and can be  compared to the mass-based concentration estimates from TNG. Overall, both morphological indicators agree reasonably well with observations over two orders of magnitude in stellar mass ($M_{\star}\approx10^{9.5}-10^{11.5}~\mathrm{M}_{\odot}$).}
    \label{fig:fraction_sph}
\end{figure}

\begin{figure*}
	\includegraphics[width=\textwidth]{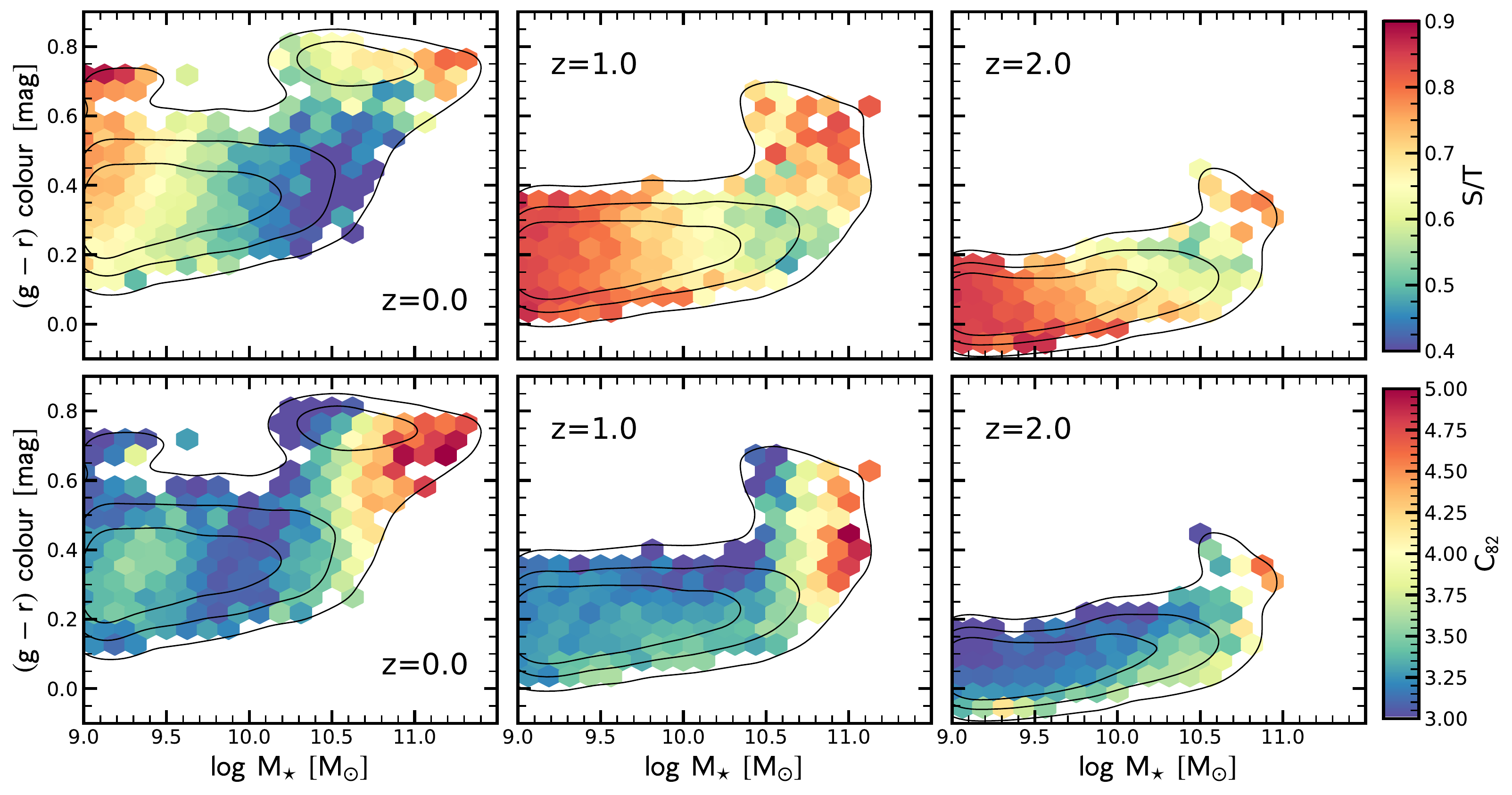}
    \caption{Relation between morphology, stellar mass, and ($g-r$) colour. We plot S/T (top) and $C_{82}$ (bottom) in the plane of ($g-r$) colour and $M_{\star}$, each bin is coloured according to its median S/T or $C_{82}$ value according to the colour bars on the right. Each bin encloses at least 10 galaxies. The black contours show the number density distribution of galaxies. The left, middle, and right panels show the galaxy population at $z=0$, $z=1$, and $z=2$, respectively. We find that $C_{82}$ depends only on $M_{\star}$, while S/T shows a $M_{\star}$ and colour dependence.}
    \label{fig:morphology_col_mass}
\end{figure*}

Observationally, the bulge-to-total ratio increases with galaxy mass in both star-forming and quiescent galaxies. Only the most massive (and quiescent) galaxies are dominated by the spheroidal component \citep[e.g.,][]{cappellari16}. At lower masses ($M_{\star}\approx10^9~\mathrm{M}_{\odot}$), it is not clear how well photometric bulge-disc decompositions agree with kinematical approaches. Figure~\ref{fig:fraction_sph} demonstrates that S/T and $C_{82}$ in IllustrisTNG roughly agree with these trends at $z=0$. 

We begin by considering our kinematic spheroid-disc decompositions, S/T, in the top panel of Figure~\ref{fig:fraction_sph}. As shown in Figure~\ref{fig:morphology_vs_M}, S/T has a minimum at intermediate masses and increases towards both low and high masses. We compare this trend to two observational samples. First, \citet{zhu18} present stellar orbit distributions for a sample of 300 nearby galaxies from the CALIFA survey \citep{sanchez12}. Derived from stellar kinematic maps via orbit-based modelling \citep{schwarzschild79}, these orbits allow them to determine the fraction of stellar mass that is in the kinematically cold, warm, hot and counter-rotating components. We estimate the spheroid component by subtracting from the total mass 1.5 times the cold component in order to take into account the different definitions of the cold/disc component in their and our works. Although their sample represents the largest such analysis to date, the sample is incomplete at low masses towards $10^9~\mathrm{M}_{\odot}$.

Overall, we find good agreement between the \citet{zhu18} observations and IllustrisTNG. They detect a smooth increase in the dynamically hot component with stellar mass above $10^{10}~\mathrm{M}_{\odot}$, which matches the stellar-mass trend in IllustrisTNG. At $10^9-10^{10}~\mathrm{M}_{\odot}$, however, the kinematically hot component also increases towards lower $M_{\star}$, indicating that these low-mass galaxies are dominated by random motion. Correspondingly, S/T increases to $\sim0.7$, implying that about 70\% of the stellar mass is a kinematically warm/hot component. This trend is in good agreement with IllustrisTNG morphologies. Consistent with this, \citet{xu19} find that the TNG100 simulation broadly reproduces the observed fractions of different orbital components and their stellar mass dependencies. 

As a second observational dataset, we consider the Galaxy and Mass Assembly (GAMA) survey \citep{driver11}, which contains more than 7500 objects at redshift $0.002<z<0.06$. \citet{moffett16} use the \citet{lange16} structural decomposition that relies on photometric data alone. They use visual morphology as a prior on whether to model a galaxy with a single or multiple components. For all galaxies where both bulge and disc components are present, they derive bulge and disc $g-i$ colours and $i$-band magnitudes, which they use to estimate component stellar masses. They then present the stellar mass fraction in ellipticals, bulges, and discs from S0--Sa, pure discs, and irregulars. The majority of low-mass objects are single-component galaxies, pure discs as well as irregular galaxies. Based on these photometric observations, we consider two different approaches to determine S/T. First, we assume that elliptical galaxies and bulges are accounting for all of the spheroidal mass. Secondly, in addition to ellipticals and bulges, we also consider irregular galaxies as spheroids. Since irregular galaxies contribute only to the low-mass regime, these two approaches differ significantly at low masses: the latter leads to an increasing S/T towards low masses, while the former leads to a strong decrease. 

The top panel of Figure~\ref{fig:fraction_sph} shows that the second method (ellipticals + bulges + irregulars) leads to much better agreement with IllustrisTNG. The overall mass dependence and slope of the relations are similar. At higher masses ($M_{\star}>10^{10}~M_{\odot}$), we find that TNG is in good agreement with the GAMA data, with more massive galaxies having a higher S/T (also consistent with findings by, e.g., \citealt{thanjavur16}). The large error-bars at the high-mass end in the observations can be understood by the small volume probed by GAMA, which implies poorer sampling of relatively rare high-mass galaxies \citep{bernardi13}. The dip at $M_{\star}=10^{10.5}-10^{11.0}~\mathrm{M}_{\odot}$ might have to do with mass-to-light ratio effects, namely with neglecting dust attenuation or outshining of the bulge by a star-forming disc component, which can cause an underestimation of both bulge $n$ values and bulge-to-disc ratios \citep{gadotti10, pastrav13, tacchella15, carollo16}. 

The bottom panel of Figure~\ref{fig:fraction_sph} compares the $C_{82}-M_{\star}$ relation in IllustrisTNG with observations. The data is again taken from the GAMA survey. Specifically, we use their published S\'{e}rsic fits in the $K-$band \citep{kelvin12} to derive the $C_{82}-M_{\star}$ relation. We select galaxies at $z=0.05-0.25$ with good S\'{e}rsic fits (quality flat $>2$). The error bars indicator the 16th and 84th percentiles. We find good agreement between the TNG simulations and observations.

In summary, we have demonstrated that IllustrisTNG galaxies generally reproduce the observed relation between morphology and stellar mass. At low masses, the estimates for S/T are reasonable, but resolution effects and modelling choices might be an issue. We discuss on this in Section~\ref{subsec:lowM_galaxies} and Appendix~\ref{app:resolution}. Furthermore, a detailed observational comparison of galaxy morphology in IllustrisTNG at $z=0$ was carried out by \citet{rodriguez-gomez19} and \citet{huertas-company19}; we discuss their results in Section~\ref{subsec:outstanding}.

\subsection{Morphology, star formation, and stellar mass}
\label{subsec:morph_SFR_M}

\begin{figure}
	\includegraphics[width=\columnwidth]{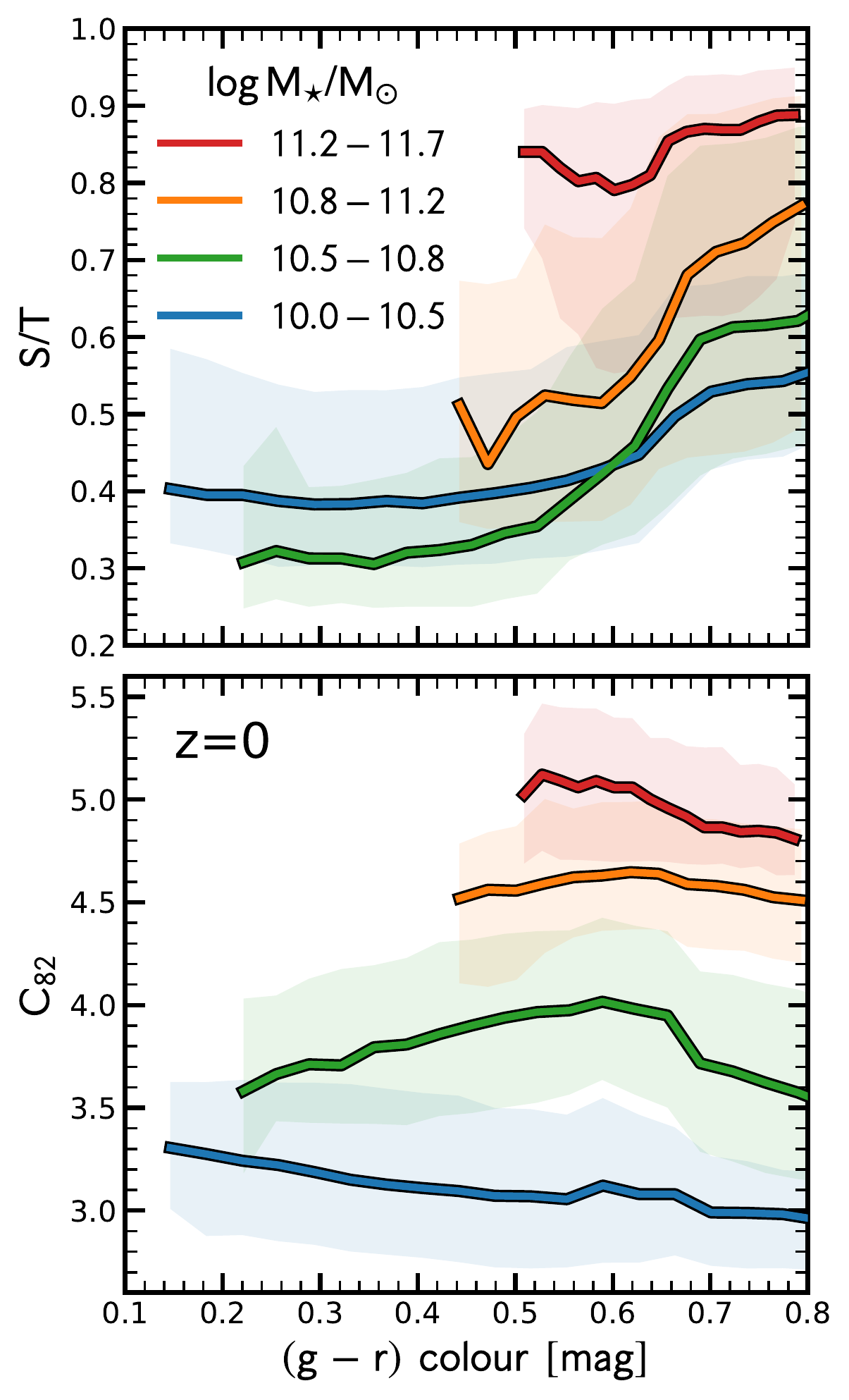}
    \caption{Morphology versus colour at $z=0$. We plot the median S/T (top panel) and $C_{82}$ (bottom panel) as a function of $(g-r)$ colour for four different mass bins. The red, orange, green and blue lines indicate the mass bins of $\log~M_{\star}/\mathrm{M}_{\odot}=11.2-11.7$, $10.8-11.2$, $10.5-10.8$, and $10.0-10.5$, respectively. We find that S/T increases with redder colours, in contrast with concentration, which barely depends on colour at fixed stellar mass. As this figure considers only $z=0$, it does not imply that galaxies necessarily need to change their morphology (kinematics) as they are ceasing their star formation and redden.}
    \label{fig:morphology_color}
\end{figure}

Figure~\ref{fig:morphology_col_mass} shows the galaxy population in the plane of $(g-r)-M_{\star}$ at redshifts $z=0$, 1, and 2 (see Appendix~\ref{app:MS_morph} for the same figure in SFR$-M_{\star}$ space). The black contours indicate the number density of IllustrisTNG galaxies, which clearly form a blue cloud and a red sequence that gets less populated towards higher redshifts. The overall colour-mass distribution is in good agreement with SDSS data \citep{nelson18_color}. Galaxies in IllustrisTNG typically cease their star formation because of black-hole feedback, which becomes effective above a black-hole mass of $M_{\rm BH}\approx10^{8.2}~\mathrm{M}_{\odot}$, i.e., when the black hole preferentially resides in the low-accretion mode and feedback is momentum-driven (``kinetic mode''; \citealt{nelson18_color,weinberger18}). Although there is a tight relation between SFR and $M_{\rm BH}$ (see Terrazas et al., in preparation), we find that galaxies transition to quiescence over a wide range in stellar mass, which can be explained, at least in part, by the scatter in the $M_{\star}-M_{\rm BH}$ relation. 

The colour coding in Figure~\ref{fig:morphology_col_mass} corresponds to S/T for the upper panels and to $C_{82}$ for the lower panels. As discussed in the previous two sections and shown in Figure~\ref{fig:morphology_vs_M}, both indicators exhibit a strong trend with stellar mass. However, in Figure~\ref{fig:morphology_col_mass}, it appears that, at fixed $M_{\star}$, S/T correlates significantly with colour, and more so the lower the redshift and above $10^{10}~M_{\odot}$, while $C_{82}$ does not. Namely, above $\sim10^{10-10.5}~\mathrm{M}_{\odot}$, redder galaxies have larger spheroidal mass fractions, by up to 0.3 in S/T values. To clarify this colour-morphology relation, Figure~\ref{fig:morphology_color} shows the median relation between colour and S/T or $C_{82}$ for four different mass bins. We confirm that S/T increases significantly with redder colour, meaning that redder galaxies have a larger spheroidal component. This is especially true for galaxies in the transition region with $M_{\star}\approx10^{10.5}-10^{11}~\mathrm{M}_{\odot}$. Conversely, $C_{82}$ remains roughly constant with colour, in particular at $z=0$ \citep[see][for a similar result]{rodriguez-gomez19}. At earlier cosmic times ($z=1$ and $z=2$), we find a secondary, weaker trend within the blue cloud: galaxies with bluer colours (higher sSFRs, see Figure~\ref{fig:MS_morph}) have slightly higher concentrations than galaxies with redder colours.

From Figure~\ref{fig:morphology_color}, it is tempting to conclude that quenching galaxies increase their S/T while not changing $C_{82}$. However, this conclusion cannot be derived from the figure because it shows the galaxy population at a specific moment in time. Specifically, today's red galaxies ceased their star formation over a wide range of cosmic times. Their star-forming progenitors also evolved significantly with time, which could lead to an observed correlation between star-forming and quiescent galaxies at a given epoch \citep[e.g.,][]{van-dokkum96, saglia10, carollo13a, lilly16, fagioli16, tacchella17_S1}. 

In the upcoming sections, we will follow galaxies through cosmic time and explicitly show how, and if, their morphological estimators change as they transition through the green valley and after quenching.  Given our findings from Figure~\ref{fig:morphology_vs_M}, we can already speculate that a reason for the apparent correlation between S/T and colour in Figure~\ref{fig:morphology_color} is that galaxies that have ceased their star formation in the past had a higher S/T value on average.

\subsection{The build-up of the different stellar components}

\begin{figure*}
	\includegraphics[width=\textwidth]{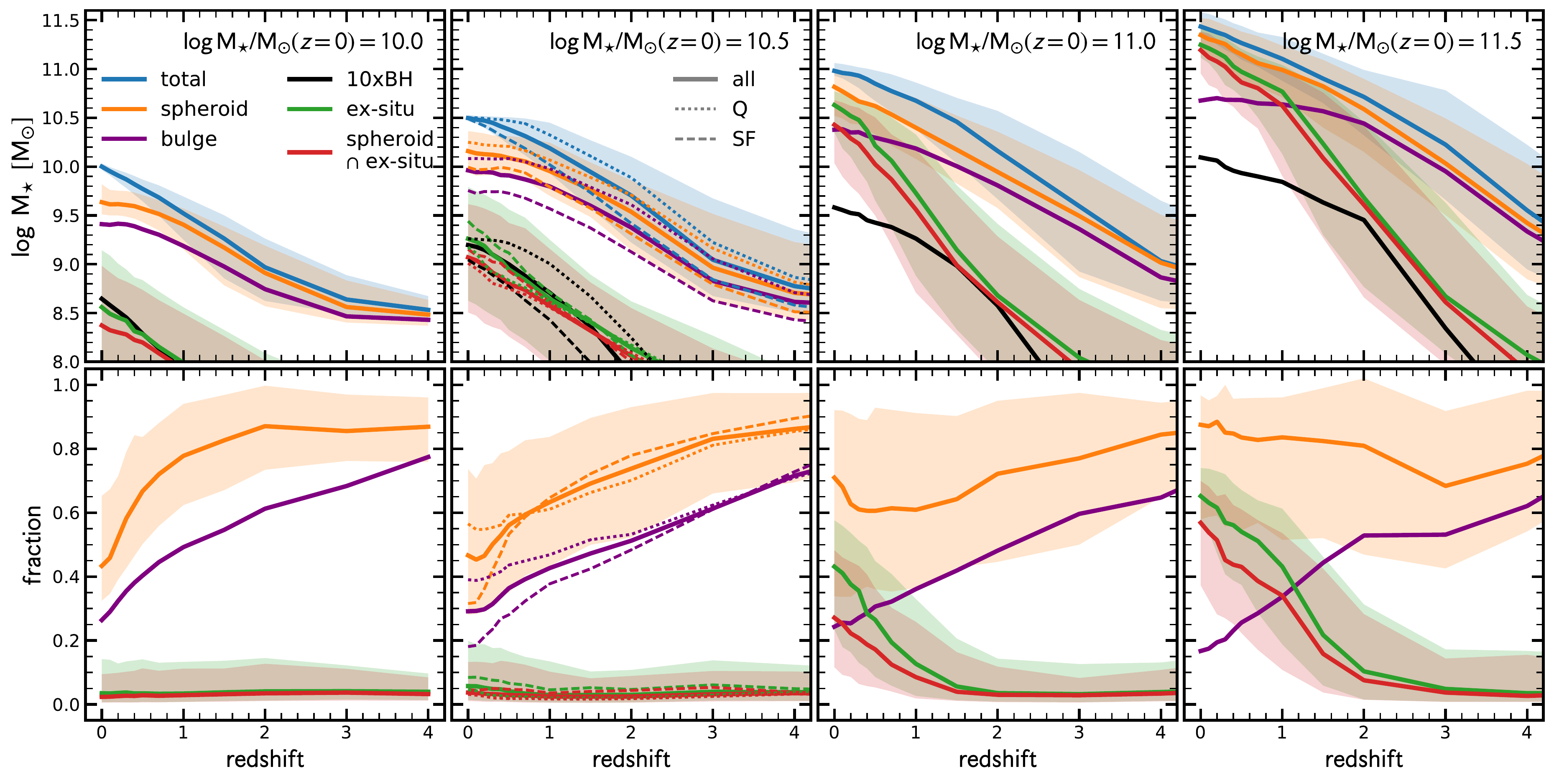}
    \caption{Evolution of the different stellar components following the main progenitor galaxy as a function of redshift. The blue, orange, green, red, purple and black lines correspond to the total stellar mass, spheroidal component, ex-situ component, spheroid\&ex-situ component, bulge component (spheroidal mass within central 3 kpc), and black-hole mass (multiplied by a factor of 10), respectively. The panels from the left to the right show bins of increasing stellar mass. We plot star-forming (dashed) and quiescent (dotted) galaxies separately in the mass bin $M_{\star}\approx10^{10.5}~\mathrm{M}_{\odot}$, since only this mass bin has a significant fraction of both star-forming and quiescent galaxies ($>30\%$ in each category). The top panels show the mass evolution in absolute units, the bottom panels show the fraction contributed by the different components with respect to the total stellar mass. For intermediate-mass galaxies ($M_{\star}\approx10^{10.0}-10^{10.5}~\mathrm{M}_{\odot}$), the mass in the spheroidal component dominates at early times ($z\ga2$) but decreases significantly in recent times: these galaxies have grown a disc component. Higher mass galaxies ($M_{\star}\approx10^{11.0}-10^{11.5}~\mathrm{M}_{\odot}$) have a rather constant spheroid mass fraction with cosmic time, while their ex-situ (and spheroid\&ex-situ) mass fraction strongly increases since $z\approx1-2$, meaning that they grow mostly through mergers.}
    \label{fig:mass_budget_evolution}
\end{figure*}

In this section, we investigate how the different components of stellar mass assemble. We select galaxies in four stellar-mass bins at $z=0$: $10^{10}~\mathrm{M}_{\odot}$ (with a bin width of $\pm0.05~\mathrm{dex}$), $10^{10.5}~\mathrm{M}_{\odot}$ ($\pm0.05~\mathrm{dex}$), $10^{11.0}~\mathrm{M}_{\odot}$ ($\pm0.1~\mathrm{dex}$), and $10^{11.5}~\mathrm{M}_{\odot}$ ($\pm0.2~\mathrm{dex}$). Each bin contains roughly $100-200$ galaxies. We now follow the progenitors of each galaxy and track the evolution of their components. Figure~\ref{fig:mass_budget_evolution} shows this evolution in total stellar mass, the spheroidal component, the ex-situ component, the spheroid\&ex-situ component (i.e., stars that have formed ex-situ and are part of the spheroidal component), the bulge component (spheroidal component within the inner 3 kpc), and the black hole mass (multiplied by $10$ so that it can be shown on the same scale). We separately consider star-forming (dashed) and quiescent (dotted) galaxies for the $M_{\star}\approx10^{10.5}~\mathrm{M}_{\odot}$ mass bin, since only this bin has a significant fraction of both star-forming and quiescent galaxies ($>30\%$ in each category). The lower mass bin is dominated by star-forming galaxies, the higher mass bins consist mainly of quiescent galaxies.

Today's galaxies with $M_{\star}\approx10^{10}~\mathrm{M}_{\odot}$ are typically dominated by their own star formation, with a negligible ex-situ contribution at all cosmic times \citep[see also][]{rodriguez-gomez16, gomez17, qu17, clauwens18, pillepich18_cluster}. At early times ($z\approx1-3$), the spheroidal component grows hand-in-hand with the total stellar mass while the bulge component grows at a slightly slower pace, resulting in a decreasing bulge fraction. Thus, most of the mass growth of the spheroidal component is associated with growth in the outer regions ($>3~\mathrm{kpc}$). During this early time, the galaxies are dominated by random stellar motions. In recent times ($z=0-1$), they evolve towards a disc-dominated morphology: the spheroidal and bulge components grow only weakly (by about 0.2 dex) and their fractional contributions drop to $\sim45\%$ and $\sim25\%$, respectively. This decrease demonstrates that those galaxies go through a phase of disc formation.

Galaxies with $M_{\star}\approx10^{10.5}~\mathrm{M}_{\odot}$ at $z=0$ follow a similar growth pattern: their spheroidal and bulge components grow at a slower pace than the total stellar mass, leading to a decrease in spheroidal and bulge mass fraction. This disc formation phase starts at $z\sim2-3$, earlier than for lower-mass galaxies, where it begins at $z\sim1-2$. The main difference between the quiescent and star-forming galaxies is that quiescent galaxies assemble earlier and have higher black hole and spheroid masses. Interestingly, the mass of the black hole ($M_{\rm BH}=10^8~\mathrm{M}_{\odot}$ at $z=0$) tracks the mass of the ex-situ mass fraction quite closely. This points to mergers driving at least part of the black-hole mass growth, either by fuelling the central part of the galaxies with gas or simply via BH-BH mergers. Such an evolution is consistent with the black-hole growth picture of \citet[][their figure 7]{weinberger18}. This picture is also consistent with the results of Section~\ref{subsec:exsitu} and Figure~\ref{fig:fexsitu_M_ST}, where we find that a high concentration of the stellar mass is associated with a high ex-situ fraction.

Galaxies with $M_{\star}\approx10^{11}~\mathrm{M}_{\odot}$ at $z=0$ assemble rather quickly at $z>1$. Their spheroidal component grows slightly slower than their total stellar mass, giving rise to a weak decline of S/T from $\sim80\%$ at $z\sim4$ to $\sim60\%$ at $z\sim1$. However, the scatter in the population is large because some galaxies have ceased their star formation earlier than others. In an upcoming paper, we demonstrate that more recently quenched galaxies have a lower S/T value than galaxies that quench at earlier epochs. Here, we understand the weak decline as evidence that galaxies in this mass bin have not been able to efficiently form a disc component (see next section). At late time ($z<1$), the rise of the spheroidal component (and the negligible growth of the bulge) are driven by ex-situ accretion in the outskirts. The black-hole growth tracks the growth of the ex-situ component until the black hole reaches a mass of $\sim10^{8.2}~\mathrm{M}_{\odot}$. At this mass scale, the galaxy quenches because it enters the more efficient black-hole feedback mode (Section~\ref{subsec:morph_SFR_M}). 

Finally, galaxies in the most massive bin ($M_{\star}\approx10^{11.5}~\mathrm{M}_{\odot}$ at $z=0$) show a weakly increasing spheroid fraction with cosmic time. After these galaxies cease their star formation, they grow by ex-situ mass accretion, increasing their ex-situ fraction to more than 60\% at $z=0$. The spheroidal component grows at a similar pace, implying that a large fraction of the ex-situ accreted stars are deposited in the spheroid component.

\subsection{When and where do discs form?}

\begin{figure*}
  \centering
  \subfloat{\includegraphics[width=0.488\textwidth]{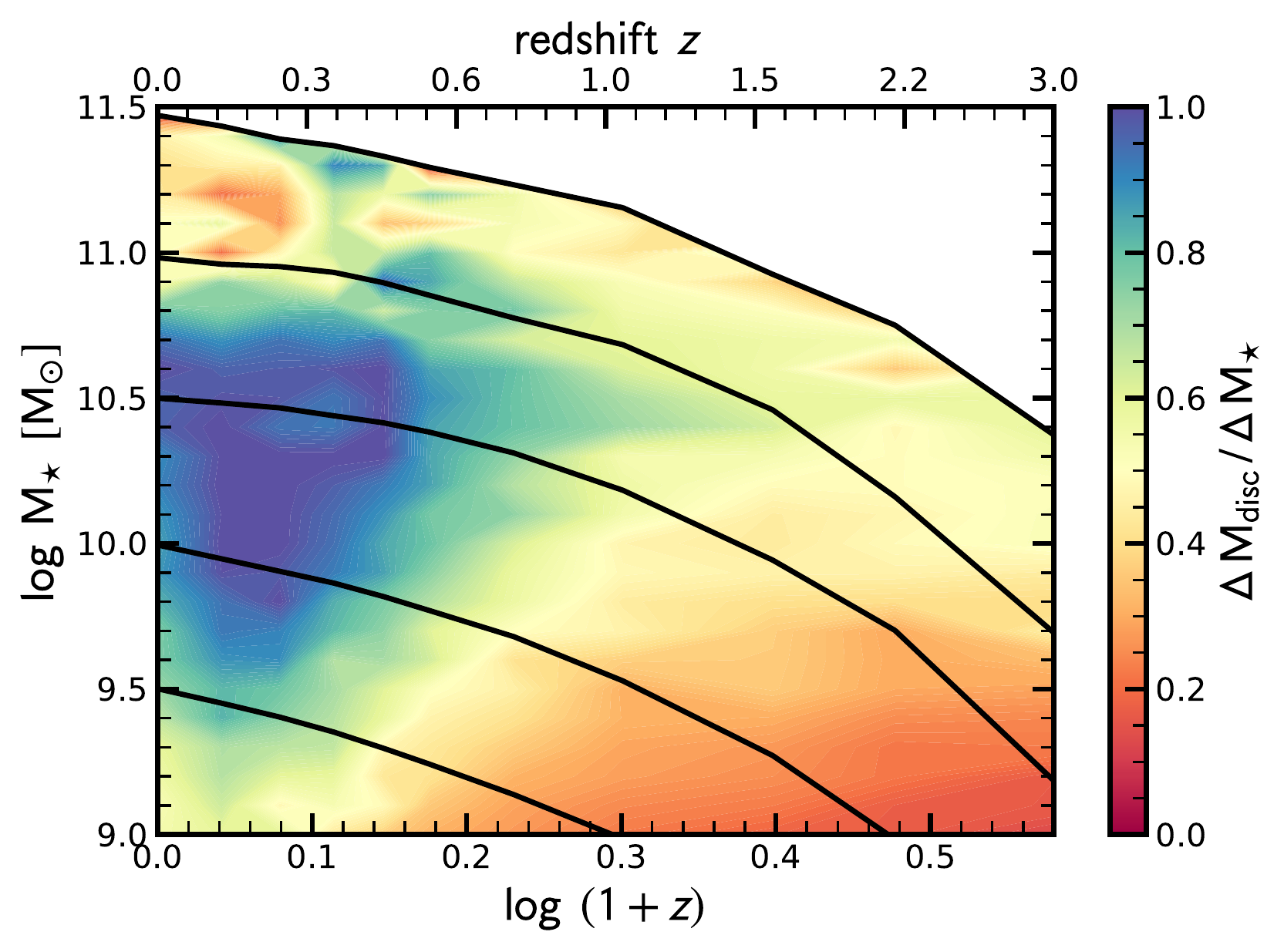}}
  \hfill
  \subfloat{\includegraphics[width=0.5\textwidth]{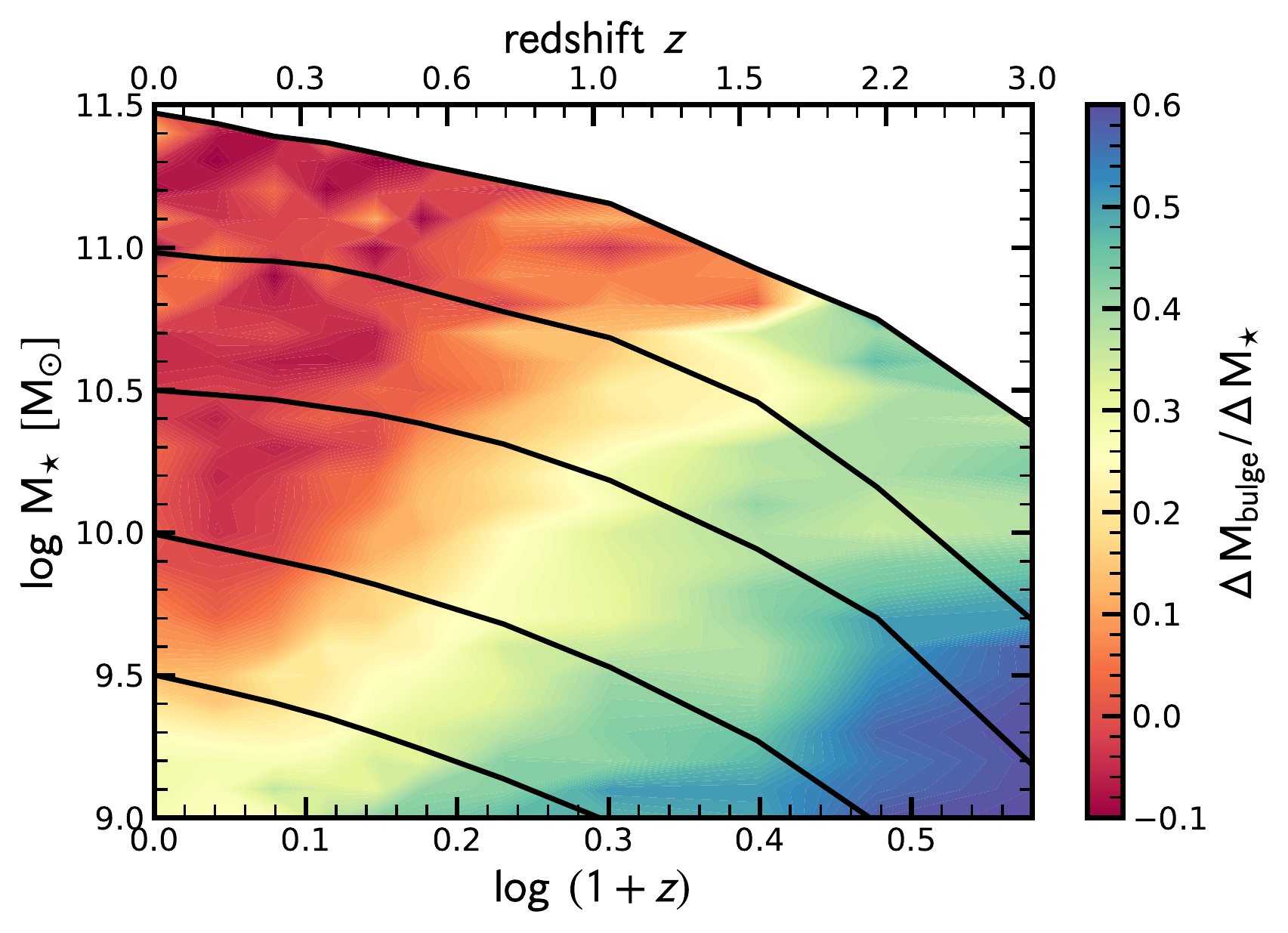}}
  \caption{Disc (left) and bulge (right) formation efficiency as a function of stellar mass and redshift. The colour coding indicates the fractional mass increase in the disc and bulge component relative to the change in the total stellar mass. The bulge mass is defined as the spheroidal mass within the inner 3 kpc. The solid black lines show the median mass growth histories of galaxies in bins of $z=0$ stellar mass. At high redshifts ($z>2$) and low masses ($M_{\star}\approx10^{9-10}~\mathrm{M}_{\odot}$), the formation of significant disc components is inefficient, while bulges form efficiently. The disc formation efficiency increases with cosmic time, in particular for stellar masses of $10^{9.5}-10^{10.5}~\mathrm{M}_{\odot}$, with lower masses entering the efficient regime at later times. At the highest masses, disc formation is always suppressed because at late times those galaxies grow mainly through mergers, leading to spheroid assembly in the outskirts.}
  \label{fig:disk_growth}
\end{figure*}


We now turn to the efficiency of disc and bulge formation as a function of stellar mass and redshift. Do all galaxies form discs at a certain efficiency, or is there a particular time and mass range where discs form?  We define the disc (bulge) formation efficiency as the fraction of added stellar material that enters into ordered rotation (random motion within the inner 3 kpc). Figure~\ref{fig:disk_growth} presents the disc (left) and bulge (right) formation efficiency, computed by following $z=0$ galaxies back in time and measuring the change of their disc and bulge mass relative to the change of their total mass. Here, the disc is defined as rotationally supported stars with $j_z/j > 0.7$, while the bulge is defined as the spheroidal component within the inner 3 kpc. The black lines in Figure~\ref{fig:disk_growth} show the median mass growth histories of galaxies in different mass bins. 

Figure~\ref{fig:disk_growth} highlights a key result: the efficiency of disc formation, and thus the build-up of morphology, depends strongly on both $M_{\star}$ and cosmic time. At early times, $z>2$, the formation efficiency of discs remains low, in particular for galaxies with $M_{\star}<10^{10}~\mathrm{M_{\odot}}$. With passing cosmic time, the formation efficiency of the disc component increases, in particular for intermediate-mass galaxies with $M_{\star}\approx10^{10.5}~\mathrm{M}_{\odot}$. The lowest mass galaxies with $M_{\star}\la10^{9.5}~\mathrm{M}_{\odot}$ have only very recently entered the region of efficient disc formation, giving rise to their high S/T ratios (Section~\ref{subsec:mass_dependence}). Similarly, massive galaxies with $M_{\star}\ga10^{11}~\mathrm{M}_{\odot}$ have never been able to efficiently form a disc. At early times, they quickly formed stars but their disc formation efficiency was, at best, intermediate. At late times, their growth is dominated by ex-situ mass accretion that mostly deposits stars in the spheroid component. The ex-situ accretion can also destroy discs, though this mechanism does not dominate. This can be seen from Figure~\ref{fig:disk_growth}, where the median $\Delta\mathrm{M}_{\rm disc}/\Delta\mathrm{M}_{\star}$ is always positive.

Focusing on the assembly of bulges, i.e. the spheroidal component in the inner 3 kpc, we find to first order that bulges form efficiently when disc growth is inefficient. An exception is the regime of massive galaxies at $z<1$, where disc and bulge formation are both inefficient. This is because the growth via mergers leads to growth of the spheroid component in the outskirts, and leaves the spheroidal mass within the inner 3 kpc nearly unaffected (second-order effects such as mass-loss and adiabatic expansion can lead to a decrease in the bulge component). The key point to take away is that bulges form efficiently at early cosmic times when galaxies were star forming. We explore this further in the next section.

\subsection{Spheroids assemble while galaxies are star forming}

\begin{figure}
	\includegraphics[width=\columnwidth]{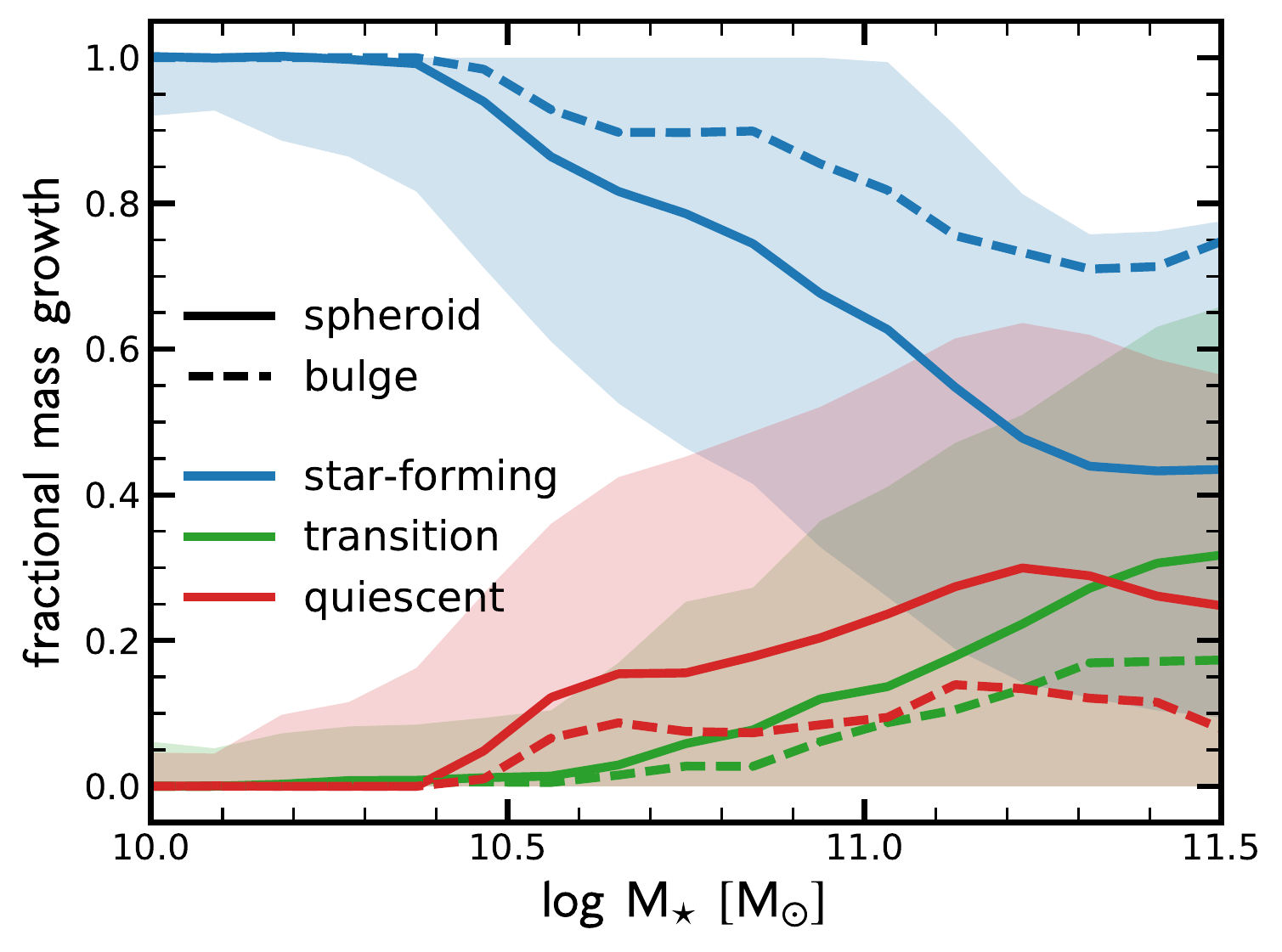}
    \caption{The fraction of the spheroidal mass of $z=0$ galaxies that has been assembled while the galaxy was star-forming (blue), transitional (green), and quiescent (red). At $M_{\star}\la10^{10.5}~\mathrm{M}_{\odot}$, most central galaxies have always been star-forming, meaning that the spheroid component has to have assembled during this phase. Towards higher masses, the fractional contribution of the star-forming phase decreases continuously, while the contribution of the transitional and quiescent phases increases. At the highest masses (towards $10^{11.5}~\mathrm{M}_{\odot}$), about half of the spheroid component has been assembled in the star-forming phase while the other half originated in the transition and quiescent phases. The shaded regions indicate the 16-84th percentile, highlighting the large diversity of galaxy histories at fixed $M_{\star}$.}
    \label{fig:fraction_growth_Msph}
\end{figure}

\begin{figure*}
	\includegraphics[width=\textwidth]{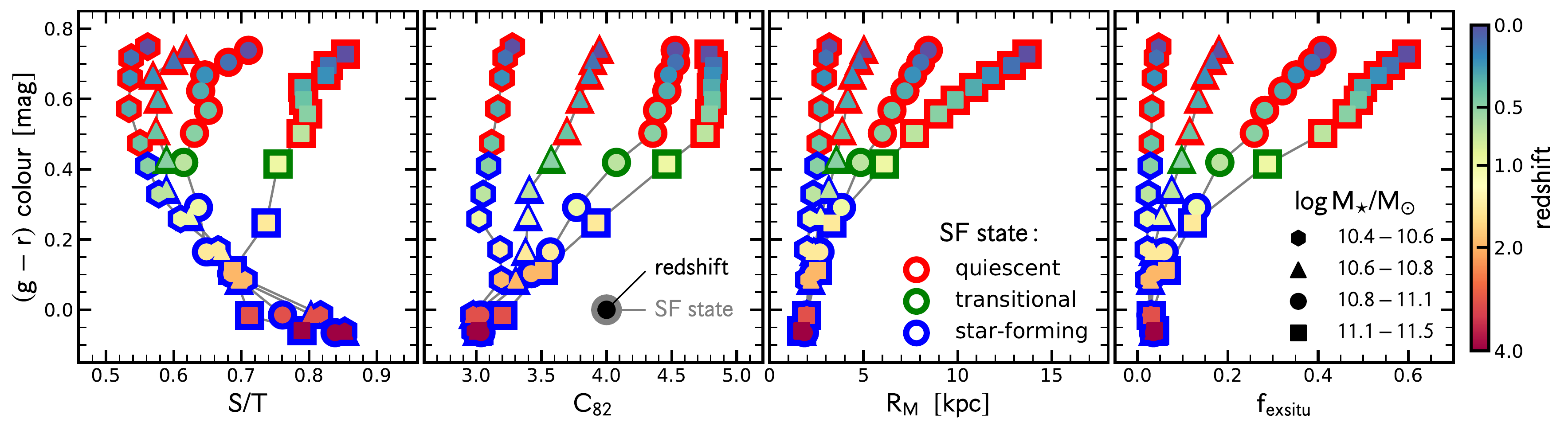}
    \caption{Median evolutionary tracks for the morphological properties of galaxies with respect to the ($g-r$) colour. The four tracks in each panel correspond to quiescent galaxy samples with $M_{\star}=10^{10.4}-10^{10.6}$ (hexagons; 448 galaxies), $M_{\star}=10^{10.6}-10^{10.8}$ (triangles; 348 galaxies), $M_{\star}=10^{10.8}-10^{11.1}$ (circles; 330 galaxies), and $M_{\star}=10^{11.1}-10^{11.5}$ (squares; 147 galaxies). The symbols' face colours correspond to redshift as indicated on the right. The symbols' edge colours mark whether the sample of galaxies is classified as star forming (blue), transitioning (green), or quiescent (red), according to their sSFR with respect to the Hubble time. The panels from the left to the right show the evolutionary tracks in the planes of ($g-r$) colour versus S/T, $C_{82}$, stellar half-mass size ($R_{\rm M}$), and ex-situ stellar mass fraction ($f_{\rm exsitu}$). We find that S/T decreases significantly during the star-forming phase (i.e. galaxies build their stellar discs), but does not change significantly during the process of quenching: $10^{10-11}~\mathrm{M}_{\odot}$ galaxies increase their spheroidal mass fraction only at late times, once they are already quenched. Conversely, $C_{82}$ increases just before quenching. $R_{\rm M}$ and $f_{\rm exsitu}$ also increase significantly, especially for the most massive galaxies, highlighting the importance of ex-situ accretion in the outskirts.}
    \label{fig:evolutionary_tracks}
\end{figure*}

We now turn to the question of when the spheroidal component forms: during the star-forming phase, or after galaxies have ceased their star formation? Figure~\ref{fig:fraction_growth_Msph} shows the fraction of  spheroid and bulge (spheroidal component in the inner 3 kpc) that was added during the star-forming phase (blue), the transition phase (green), and the quiescent phase (red; see Section~\ref{subsec:SF_definition} for the definitions). Here we do not follow backwards in time the stars that are in the spheroidal and bulge components of the $z=0$ galaxies. We rather evaluate the mass fractions in the two components in galaxies at different times in comparison to their $z=0$ mass and assume that stars that are on hot orbits at some point in time do remain in high-dispersion states at subsequent times. 

The key result is that, at all masses, a large fraction of the spheroid is added while galaxies are star-forming, albeit with a large scatter highlighting the diversity at a given $M_{\star}$. At low masses, ($M_{\star}<10^{10.5}~\mathrm{M}_{\odot}$), most galaxies are star-forming throughout their lifetimes. Thus, by construction, all of their spheroidal mass growth takes place while star-forming. Towards higher stellar masses, the fractional mass growth while star forming steadily decreases, while the mass growth during the transition and quiescent phases increase at roughly the same pace. However, even at the highest masses ($M_{\star}\approx10^{11.5}~\mathrm{M}_{\odot}$), $\sim45\%$ of the spheroidal mass has been added while the galaxies were star forming, while about 30\% and 25\% was added in the transition and quiescent phases, respectively. 

Focusing on the central ($<3$ kpc) bulge component, the fraction of mass assembled during the star-forming phase is even higher and always at least 80\%. The bulge growth during the quiescent phase is negligible ($<5\%$ at all masses). This finding highlights that bulges in massive galaxies have already assembled when galaxies stop forming stars, while (dry) mergers play an important role in building the spheroidal component in the outskirts. 

Clearly, the fractions shown in Figure~\ref{fig:fraction_growth_Msph} depend on the exact definition of the star-forming, transition, and quiescent phases. As outlined in Section~\ref{subsec:SF_definition}, we adopt a rather conservative cut on star-forming galaxies; a simple colour cut would classify more galaxies as star-forming. Lowering this threshold would increase the star-forming fraction further, and thus strengthen our statement that a large fraction of the spheroid formation takes place while galaxies are star-forming. However, even when switching to a simple colour cut, the trends qualitatively remain the same, highlighting that our main conclusion is robust.

\subsection{Evolution with respect to star formation and colour}
\label{subsec:evo_tracks}

We now advance the discussion to the relation between morphological changes and colours: how much of the morphological transformation takes place while galaxies are blue and star-forming versus red and quiescent? Figure~\ref{fig:evolutionary_tracks} shows the median evolutionary tracks of the colours of four samples of quiescent galaxies, marked by different symbols. The symbols' face colours correspond to redshift as indicated on the right. The symbols' edge colours mark whether the sample of galaxies is classified as star forming (blue), transitioning (green), or quiescent (red), according to their sSFR with respect to the Hubble time (Section~\ref{subsec:SF_definition}). Although individual galaxies need not follow these median trends, they serve as a rough indication for the typical relation between colour and morphological changes. Specifically, we compare the colour evolution of quiescent galaxies in four stellar mass bins to the evolution of their spheroid-to-total ratio (S/T), concentration ($C_{82}$), stellar half-mass size ($R_{\rm M}$), and ex-situ stellar mass fraction ($f_{\rm exsitu}$). By construction, all median tracks evolve to red colours of $(g-r)>0.7~\mathrm{mag}$ by $z=0$. At early cosmic times ($z>2$), the colours of all samples are blue, i.e. dominated by young stellar populations and high specific SFRs. The colour of the most massive quiescent galaxies evolves most rapidly, followed by the intermediate-mass objects.

Focusing on the evolution of S/T (left panel of Figure~\ref{fig:evolutionary_tracks}), we find that all but the most massive galaxies are rapidly building up their discs during their star-formation phase (moving towards lower S/T). Once star formation has ceased, the median tracks evolve at almost constant S/T for a time. This non-evolution demonstrates that galaxies do not (or at least not coherently) transform their morphology as expressed by S/T during quenching or transitional phase. During the quiescent phase, i.e. when they are already quenched, galaxies' S/T increases by $0.05-0.1$, typically at $z\la0.5-0.3$ and plausibly due to mergers (see previous sections).

The concentration (second panel from the left) increases with time at $z<2$ for all four samples of galaxies while they are star-forming. The increase is slowest for the lowest mass galaxies. For the most massive galaxies, $C_{82}$ increases significantly just before they halt their star formation, i.e. during the transition phase, while there is little change in $C_{82}$ during the subsequent quiescent phase. This result demonstrates, once again, that massive galaxies build up their high stellar mass concentration during the star-formation phase.

The third panel in Figure~\ref{fig:evolutionary_tracks} shows the evolution of galaxy size as quantified by the stellar half-mass radius. The size growth of quiescent galaxies depends strongly on mass: while all galaxy samples grow their sizes at all times, that growth is much more pronounced for the most massive galaxies \citep{genel18}. The size growth in the transition and quiescent phases resembles the increase in the ex-situ stellar mass fraction (right panel), indicating that the size growth is likely due to the accretion of stellar mass onto the outskirts. However, in particular during the transition phase, star formation in the outskirts could also play an important role in the increase in the sizes of these galaxies, an effect that should be investigated further in future work. 

The evolutionary tracks in Figure~\ref{fig:evolutionary_tracks} follow quiescent galaxies back in time. Clearly, we can also investigate different sub-samples at earlier cosmic times and study their subsequent evolution. For example, focusing on quiescent disc galaxies at $z=1-2$, we find that most of those galaxies increase their S/T value by 0.2-0.3 by $z=0$, i.e. they turn into spheroids, while staying quiescent. This is consistent with Figure~\ref{fig:evolutionary_tracks}, where we see that S/T increases at late times. We will study this further in a future publication that focuses on those galaxies in particular.

Finally, we would like to stress that Figure~\ref{fig:evolutionary_tracks} shows median tracks. Studying the scatter in the tracks (not shown because of clarity) reveals that there is significant scatter in the S/T-tracks, highlighting the diversity in evolutionary paths. On the other hand, for the concentration, size and ex-situ mass fraction, the scatter in the evolutionary paths are much tighter, implying a more coherent evolution.

\section{Observational consequences for $z=0$ galaxies}
\label{sec:consequences}

After discussing how IllustrisTNG galaxy morphologies evolve with cosmic time and stellar mass, we now turn to observational consequences of this model at $z=0$. We focus on the ex-situ stellar mass fraction and its dependence on morphology, quantify the stellar mass density profiles, and discuss stellar population properties of the disc and spheroidal components.

\subsection{Correlation between ex-situ mass fraction and morphology}
\label{subsec:exsitu}

\begin{figure}
	\includegraphics[width=\columnwidth]{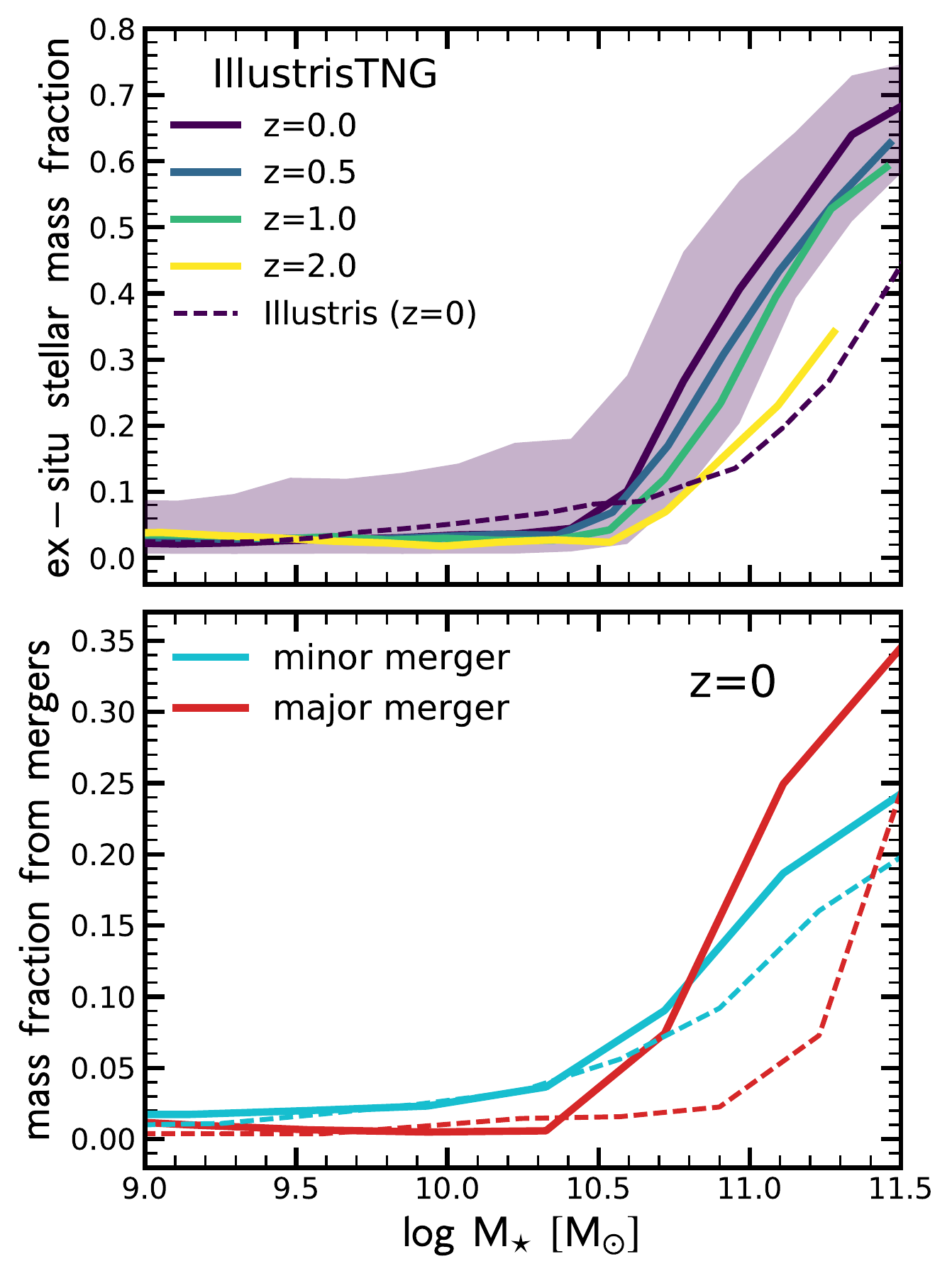}
    \caption{Ex-situ mass fraction (top) and mass fraction split by merger type (bottom) as a function of $M_{\star}$. Top panel: the solid coloured lines show the median relations of ex-situ stellar mass fraction and $M_{\star}$ at different redshifts. The dashed line shows the original Illustris simulations at $z=0$ \citep{rodriguez-gomez16}. We find ex-situ stellar mass fraction to be a strong function of $M_{\star}$: it approximately vanishes at $M_{\star}\la10^{10.5}~\mathrm{M}_{\odot}$, and then increases towards 0.7 when $M_{\star}$ approaches $10^{11.5}~\mathrm{M}_{\odot}$. Furthermore, there is a weak redshift evolution below $z\la1$ and stronger one between $z\sim2$ and 1: the ex-situ stellar mass fraction generally increases with cosmic time at a given $M_{\star}$. Bottom panel: fraction of stellar mass that was accreted from completed minor and major mergers. Major mergers are defined by a stellar mass ratio $>1:4$, while minor mergers are all mergers that are not major. The solid and dashed lines show the results for IllustrisTNG and original Illustris, respectively. Overall, the stellar mass growth above $10^{11}~\mathrm{M}_{\odot}$ is dominated by ex-situ mass accretion. The results for Illustris are qualitatively similar, but predict lower ex-situ fractions at high stellar mass. This trends can be understood based on changes in the feedback and the resulting stellar mass function.}
    \label{fig:fexsitu_M}
\end{figure}

\begin{figure}
	\includegraphics[width=\columnwidth]{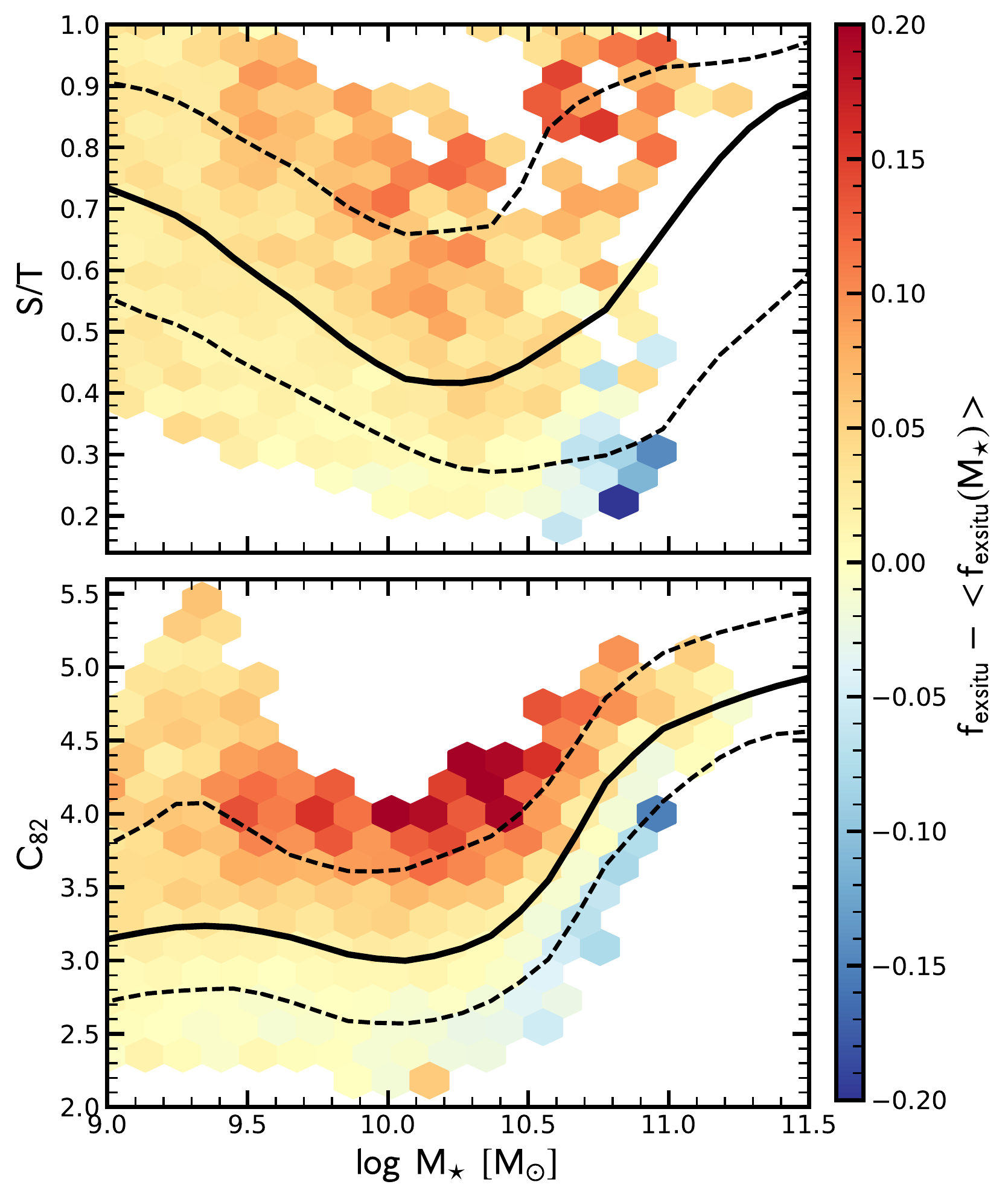}
    \caption{Normalized ex-situ mass fraction in the plane of S/T (top panel) and $C_{82}$ (bottom panel) versus $M_{\star}$. Each bin (containing at least 10 galaxies) is colour-coded by its average ex-situ stellar mass fraction after removing the $M_{\star}$-dependence of $f_{\rm exsitu}$ (Figure~\ref{fig:fexsitu_M}). The solid and dashed black lines indicate the median and 16th/84th percentiles of the S/T$-$ and $C_{82}-M_{\star}$ relations, respectively. After controlling for the $M_{\star}$-dependence, we find a secondary trend with morphology: at fixed $M_{\star}$, higher values of S/T and $C_{82}$ correlate with higher-than-average ex-situ stellar mass fractions.}
    \label{fig:fexsitu_M_ST}
\end{figure}

Many of the results in the previous sections imply that mergers play an important role in shaping galaxies. In this section, we thus investigate the relation between morphology and the in-situ and ex-situ stellar components of simulated galaxies. The ex-situ stellar mass fractions for IllustrisTNG galaxies have already been presented in \citet{pillepich18_cluster}. For completeness, we show the ex-situ fractions in Figure~\ref{fig:fexsitu_M} and compare them to the original Illustris analysis of \citet{rodriguez-gomez16}. Here all measurements are within three times the stellar half-mass radius. We then expand on previous analyses by directly connecting the ex-situ fraction to morphology. 

We define and measure the ex-situ stellar mass fraction as in \citet{rodriguez-gomez16} and refer the reader to their paper for details. Briefly, the ex-situ stellar mass fraction is the fraction of the total stellar mass of a galaxy that has been contributed by stars that formed in other galaxies and were subsequently accreted. Hence, the ex-situ fraction is a good tracer of mergers, but not a perfect one: mergers do not only add stellar mass (which would be counted as ex-situ), but they can also bring in gas and enhance the star-formation activity, which would then be counted as in-situ. Therefore, the ex-situ component is a good tracer of gas-poor (dry) mergers (see also Figure 6 in \citealt{rodriguez-gomez17}). 
The top panel of Figure~\ref{fig:fexsitu_M} shows the median fraction of ex-situ stellar mass as function of stellar mass and redshift. There is a striking mass dependence: up to $M_{\star}\approx10^{10.5}~\mathrm{M}_{\odot}$, the median ex-situ fraction is $<5\%$, while at and above $10^{11}~\mathrm{M}_{\odot}$ it amounts to more than 50\%. This strong increase occurs mainly because galaxies quench efficiently and subsequently grow through ex-situ accretion. In addition, the merger rate increases with mass \citep{rodriguez-gomez15}. Moreover, the majority of stars stems from mergers with a halo mass ratio $<10:1$ \citep{genel10}, which for galaxies with $M_{\star}\la10^{10.5}~\mathrm{M}_{\odot}$ means mergers with haloes that host very few stars. Figure~\ref{fig:fexsitu_M} also shows a weak redshift evolution after $z<1$ while a non negligible one between $z\sim1$ and 2: at fixed stellar mass, the ex-situ stellar mass fraction increases with cosmic time, for example at $M_{\star}\approx10^{11}~\mathrm{M}_{\odot}$ where it increases from $\sim20\%$ to $\sim50\%$ from $z=2$ to $z=0$. This shows that the contribution from mergers and accretion becomes relevant over the last 8 Gyr or so.

The bottom panel of Figure~\ref{fig:fexsitu_M} shows the fraction of stellar mass that was accreted from completed minor and major mergers. Major mergers are defined by a mass ratio of $>1:4$. The stellar mass fractions from minor and major mergers do not exactly add up to the ex-situ stellar mass fraction (shown in the top panel), because the latter also includes the stellar mass from on-going mergers, from flybys, and from mass formed outside the galaxies. For the exact definition, see \citet{rodriguez-gomez15} and \citet{rodriguez-gomez16}. The key result is that minor mergers dominate up to a stellar mass of $M_{\star}\approx10^{10.8}~\mathrm{M}_{\odot}$, beyond which major mergers start to dominate. As we will show in the next section (Section~\ref{subsec:profile}), this translates into a different radial distribution of the ex-situ stellar mass fraction within galaxies.

The most striking conclusion from Figure~\ref{fig:fexsitu_M} is that the ex-situ fraction steeply increases with stellar mass above about $M_{\star}\approx10^{10.5}~\mathrm{M}_{\odot}$, reminiscent of the increase in S/T and $C_{82}$ seen in Figure~\ref{fig:morphology_vs_M}. However, this similarity does not prove that a high ex-situ mass fraction implies a high S/T or $C_{82}$ value; they could be independent of each other but both correlate with $M_{\star}$. To investigate the connection directly, Figure~\ref{fig:fexsitu_M_ST} shows 2D histograms of S/T and $C_{82}$ as a function of $M_{\star}$, where each bin is colour-coded by the normalized ex-situ fraction (ex-situ fraction after subtracting the median $f_{\rm exsitu}-M_{\star}$ relation to remove the strong mass dependence). At a given $M_{\star}$, galaxies with higher S/T and higher $C_{82}$ also have a higher ex-situ stellar mass fraction. This trend is particularly pronounced in intermediate-mass galaxies around $M_{\star}\approx10^{10}~\mathrm{M}_{\odot}$, which are typically star-forming. Physically, this implies that star-forming galaxies that recently experienced a merger have, on average, an increased S/T and $C_{82}$. This interpretation is consistent with a picture where mergers fuel gas to the central region to build up a spheroidal component in the core of galaxies \citep{hernquist89, barnes91, mihos94, mihos96, barnes96, tacchella16_profile}. We will investigate this scenario further in an upcoming publication, where we will study the spatially resolved SFR and $M_{\star}$ profiles.

Previously, \citet{rodriguez-gomez17} used the original Illustris simulation to study the relationship between the ex-situ fraction and kinematic morphology at $z=0$. The latter was quantified using the $\kappa_{\rm rot}$-parameter \citep[][see also Appendix~\ref{app:diff_morph}]{sales12}, which measures the fraction of kinetic energy invested in ordered rotation. Their main finding was that the correlation between the ex-situ fraction and kinematic morphology is mass-dependent: dry mergers play an important role in shaping the morphology of massive galaxies ($M_{\star}>10^{11}~\mathrm{M}_{\odot}$), while the role of mergers becomes less clear at lower masses (their Figure 5). This finding is qualitatively consistent with the upper panel from Figure~\ref{fig:fexsitu_M_ST}, which also compares the ex-situ fraction with kinematic morphology. The only quantitative difference between the results of \citet{rodriguez-gomez17} and this work is that mergers become relevant at a somewhat lower stellar mass ($10^{10}~\mathrm{M}_{\odot}$ instead of $10^{11}~\mathrm{M}_{\odot}$) in IllustrisTNG compared to Illustris, as a result of the lower transition mass between in-situ and ex-situ dominated systems in IllustrisTNG (Figure~\ref{fig:fexsitu_M}). Specifically, in Figure~\ref{fig:fexsitu_M}, dashed curves denote results from the original Illustris simulation: it predicts a lower ex-situ fraction at fixed galaxy mass towards the high-mass end in comparison to TNG (up to 25-30 percentage points). This difference between TNG and Illustris can be understood based on changes in the feedback and the resulting stellar mass function. However, it is exacerbated by a shift towards larger stellar masses of Illustris galaxies at fixed halo masses. The ex-situ fractions of TNG and Illustris galaxies differ by 10-15 percentage points at most when measured at fixed halo mass.

Similarly, \citet{bignone17} and \citet{snyder19} studied the relationship between several non-parametric morphological indicators -- including the light concentration -- and the frequency of galaxy mergers in Illustris, finding a clear correlation between the two. In particular, \citet{snyder19} found that the concentration statistic has a high importance as an indicator of very recent mergers (those that happened during the last 250 Myr; see their Figure~\ref{fig:fexsitu_M_ST}), which could be an indication of highly concentrated starbursts produced by gas-rich mergers. This finding is qualitatively consistent with the lower panel of Figure~\ref{fig:fexsitu_M_ST}, which shows a particularly strong correlation between the ex-situ fraction and concentration for intermediate-mass galaxies.

\subsection{Stellar mass density profiles}
\label{subsec:profile}

\begin{figure*}
	\includegraphics[width=\textwidth]{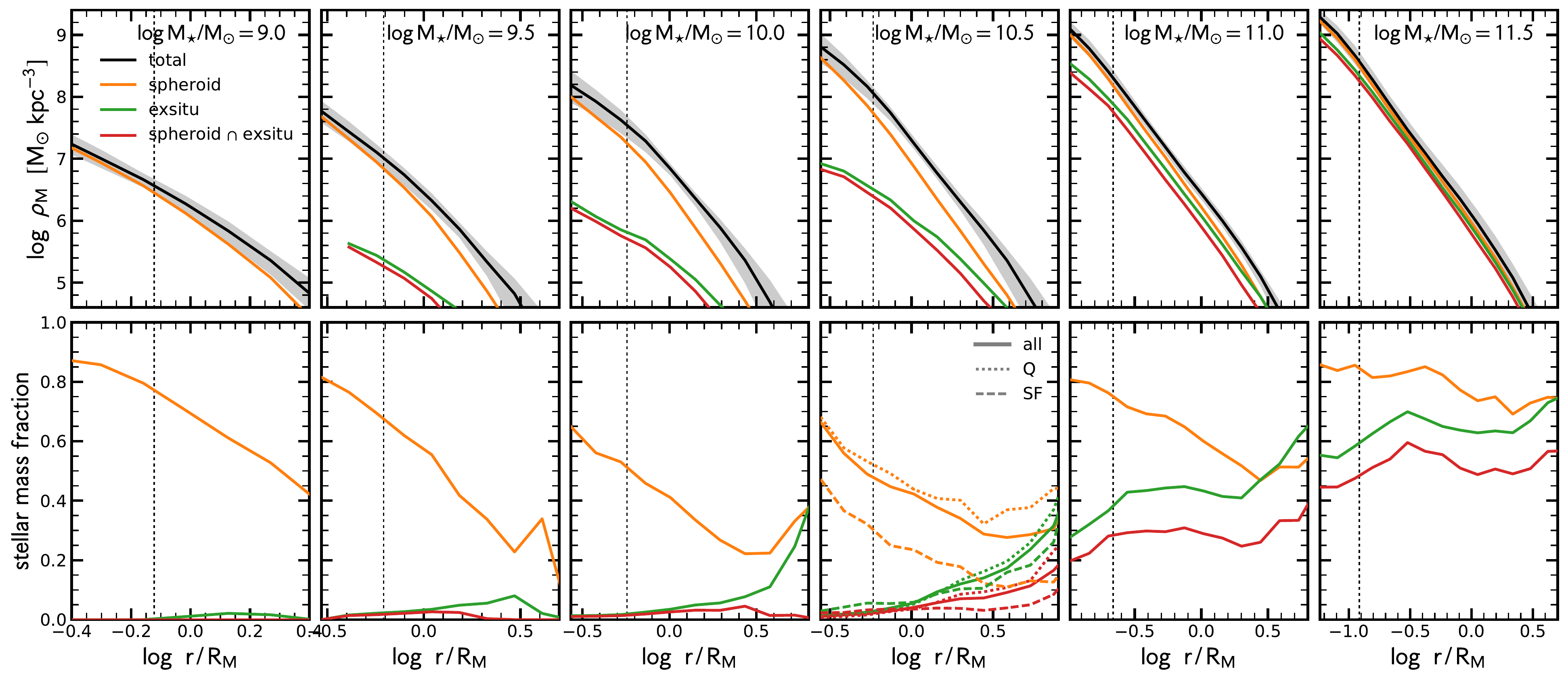}
    \caption{Stellar mass density profiles. The panels from the left to the right show increasing stellar mass bins from $M_{\star}\approx10^{9}~\mathrm{M}_{\odot}$ to $10^{11.5}~\mathrm{M}_{\odot}$. The top panels show the 3D stellar mass density profile: the black, orange, green and red lines indicate the profiles for the total stellar mass, the spheroidal component, the ex-situ component, and the spheroid\&exsitu component, respectively. The radial coordinate of all profiles are normalized by the stellar half-mass radius. The bottom panels show the fractional contributions of each component to the total mass profile. In the mass bin $M_{\star}\approx10^{10.5}~\mathrm{M}_{\odot}$ we separately show star-forming (dashed lines) and quiescent (dotted lines) galaxies, since this is the only mass bin where both populations exist in significant numbers (at least $30\%$ in each category). The ex-situ mass fraction is negligible for low-mass galaxies, and starts dominating in the outskirts of more massive galaxies. For the most massive galaxies ($M_{\star}\approx10^{11.5}~\mathrm{M}_{\odot}$), the ex-situ stellar mass fraction is roughly constant as a function of radius. The spheroidal component dominates the central part of all samples, particularly of low-mass galaxies. For the most massive galaxies, the spheroidal component is roughly constant as a function of radius. A similar figure for the original Illustris simulations can be found in Appendix~\ref{app:illustris}.}
    \label{fig:profiles}
\end{figure*}

To shed further light on the assembly of the spheroidal and ex-situ stellar mass components on spatially resolved scales, we now investigate the stellar mass distribution within galaxies at $z=0$. Figure~\ref{fig:profiles} shows the 3D stellar mass density profiles (upper panels) and the radial stellar mass fraction (bottom panels) for galaxies in six different mass bins. We split the total stellar mass profiles into the spheroidal component, the ex-situ component, and the spheroidal\&ex-situ component. The radial coordinate of the stellar mass profiles are all normalized by the stellar half-mass radii. The mass bins include both star-forming and quiescent galaxies and increase from $10^9$ to $10^{11.5}~\mathrm{M}_{\odot}$. As in Figure~\ref{fig:mass_budget_evolution}, we separate star-forming and quiescent galaxies in the bin with $M_{\star}\approx10^{10.5}~\mathrm{M}_{\odot}$, which contains significant populations of both star-forming and quiescent galaxies.

At low masses ($M_{\star}<10^{10}~\mathrm{M}_{\odot}$), galaxies have a significant spheroidal component (see also Figure~\ref{fig:morphology_vs_M}), which dominates the mass profile within the stellar half-mass radius and decreases steadily towards the outskirts. The ex-situ mass fraction is negligible at all radii. With increasing stellar mass ($M_{\star}=10^{9.5} \rightarrow 10^{10.5}~\mathrm{M}_{\odot}$), the spheroidal mass fraction steadily decreases within $3~R_{\rm M}$, but still peaks in the central region. In the outskirts of these galaxies, between 3 and $10~R_{\rm M}$, the mass profile of the spheroidal and ex-situ components increases with increasing $M_{\star}$, while the spheroid\&ex-situ component increases to a lesser degree. These trends demonstrate that intermediate-mass galaxies have a stellar envelope (halo component) that has partially been built up by the accretion of stars through minor mergers. This finding is consistent with Figure~\ref{fig:fexsitu_M}, where we show that the ex-situ fraction within $3\times R_{\rm M}$ is $<5\%$ for galaxies within this mass range, and that most of their mass is contributed by minor mergers. Nevertheless, the central spheroidal component, i.e., the bulge, consists mostly of stars that formed in-situ.

Considering even more massive galaxies, we find that the spheroidal and ex-situ mass fractions increase with stellar mass. The difference between $M_{\star}=10^{11}~\mathrm{M}_{\odot}$ and $10^{11.5}~\mathrm{M}_{\odot}$ is surprisingly stark: for galaxies with $M_{\star}=10^{11}~\mathrm{M}_{\odot}$, the spheroidal mass fraction decreases with radius, while the ex-situ mass fraction increases with radius. Galaxies with $M_{\star}=10^{11.5}~\mathrm{M}_{\odot}$, on the other hand, have basically flat spheroidal, ex-situ, and spheroid\&ex-situ profiles that contribute significantly at all radii. The natural interpretation of this difference is that the main formation channels of galaxies in the two mass bins differ  (see also Figure~\ref{fig:morphology_col_mass}). Minor mergers contribute to the ex-situ fraction in the outskirts of $M_{\star}=10^{11}~\mathrm{M}_{\odot}$ galaxies, but are inefficient at deploying significant amounts of mass in the central region \citep[e.g.,][]{rodriguez-gomez16, amorisco17}. For the most massive galaxies ($M_{\star}=10^{11.5}~\mathrm{M}_{\odot}$), however, the ex-situ mass fraction dominates the central region as well, indicating that more violent, major mergers must play a role. This picture is consistent with the stellar mass fraction from minor and major mergers shown in Figure~\ref{fig:fexsitu_M}, where we show that major mergers start to be significantly more important than minor mergers above $M_{\star}\approx10^{11.2}~\mathrm{M}_{\odot}$.

In Appendix~\ref{app:illustris}, we reproduce Figure~\ref{fig:profiles} for the original Illustris simulation. We discover large differences in the shape of the profiles, with less concentrated mass profiles in Illustris. Moreover, the spatial distribution of the spheroidal and ex-situ stellar mass fractions differs as well, highlighting that the details of stellar and black-hole feedback have a large impact on the morphology and assembly of galaxies.

\subsection{Stellar population properties of the spheroidal and disc components}

\begin{figure}
	\includegraphics[width=\columnwidth]{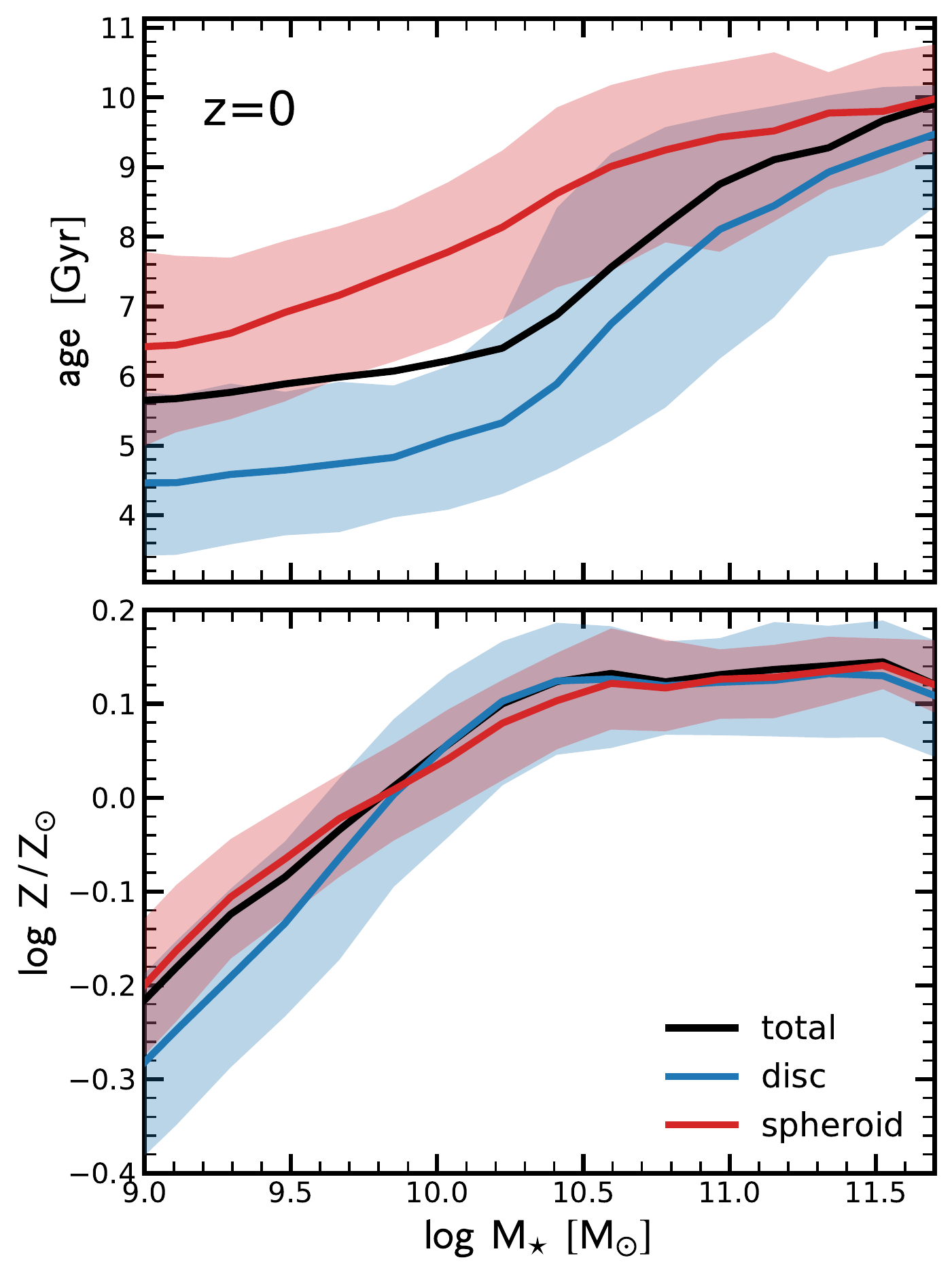}
    \caption{Stellar population properties for the spheroidal and disc components at $z=0$. The upper and lower panels show the mass-weighted stellar age and metallicity (Z) as a function of $M_{\star}$, respectively. The black, blue, and red lines show the median relations for total galaxy, the spheroidal and the disc components, respectively. The shaded regions indicate the 16th/84th percentiles. The stellar populations in more massive galaxies are older and have higher metallicity. At all masses, the spheroidal component is older than the disc component, in particular for intermediate-mass galaxies. We find no significant difference in the metallicity of the spheroidal and disc components.}
    \label{fig:SP_disk_sph}
\end{figure}

In order to gain further insight into the formation mechanism of the spheroidal and disc components, and to make predictions that can be compared to observations, we investigate the stellar population properties of the kinematic components at $z=0$. \citet{nelson18_color} presented the mass-weighted stellar age and stellar metallicity of IllustrisTNG galaxies as a function of $M_{\star}$. In Figure~\ref{fig:SP_disk_sph}, we expand on their analysis by splitting the stellar properties according to the kinematics. In particular, the top and bottom panels show the mass-weighted stellar age and stellar metallicity as a function of $M_{\star}$, with the black, blue, and red lines indicating the property of stars in the entire galaxy, in the disc component, and in the spheroidal component, respectively. 

The mass-weighted stellar age of the entire galaxy increases from 5.5 Gyr for low-mass galaxies to $>9$ Gyr for high-mass galaxies. Galaxies with $M_{\star}\la10^{10.5}~\mathrm{M}_{\odot}$ have similar ages, consistent with a nearly linear relation between SFR and $M_{\star}$ (i.e., the mass doubling timescale is roughly the same for those galaxies). The average age rapidly increases around the mass-scale where most galaxies quench ($M_{\star}\approx10^{10.5}-10^{11}~\mathrm{M}_{\odot}$). Interestingly, the spheroidal component is older than the disc component at all masses. This difference is especially pronounced in the intermediate-mass regime of $M_{\star}\approx10^{10}-10^{10.5}~\mathrm{M}_{\odot}$. This trend implies that the spheroidal component forms earlier and that the majority of the recently formed stars are in the disc component, which is consistent with disc assembly as outlined in Sections~\ref{subsec:morph_SFR_M} and \ref{subsec:evo_tracks}. 

The stellar metallicity (lower panel of Figure~\ref{fig:SP_disk_sph}) increases with $M_{\star}$ up to $M_{\star}\approx10^{10.5}~\mathrm{M}_{\odot}$ and then remains constant. Interestingly, the metallicities for the spheroidal and the disc components are practically indistinguishable. We speculate that the rather homogeneous metallicity is driven by efficient feedback, which causes metals to mix effectively. The fact that metallicity seems to correlate with the potential well can be explained by a larger stellar mass that produces more metals and by a larger halo that is able to better retain those metals \citep{marinacci14, grand18}.



\section{Discussion}
\label{sec:discussion}

We have presented a detailed analysis of how the disc and spheroidal components within galaxies evolve in the framework of the IllustrisTNG simulations. In this section, we examine the role of mergers in shaping galaxy morphology and further discuss the nature of low-mass galaxies. Finally, we highlight some remaining questions raised by our investigations, and discuss how we can improve our analysis in the future.

\subsection{The role of mergers and the emerging picture}
\label{subsec:picture}

In our standard cosmological model, dark matter haloes and their galaxies merge with each other to build larger structures. With the IllustrisTNG simulations, we confirm previous results \citep[e.g.,][]{rodriguez-gomez16, qu17, pillepich18_cluster} that the ex-situ stellar mass fraction is a steep function of stellar mass (see Figure~\ref{fig:fexsitu_M}). We put forward that there are three merger regimes: (i) at $M_{\star}\la10^{10.5}~M_{\star}$, the ex-situ fraction is small and dominated by minor mergers; (ii) at $M_{\star}\approx10^{10.5}-10^{11}~M_{\star}$, the ex-situ fraction increases strongly, and minor and major mergers are roughly equally important; and (iii) at $M_{\star}\ga10^{11}~M_{\star}$, the ex-situ fraction reaches $>50\%$ and major mergers dominate. The overall upturn of the ex-situ stellar mass fraction above $10^{10.5}~\mathrm{M}_{\odot}$ can be explained by the efficient star-formation quenching in the TNG model (primarily induced by effective black-hole feedback) around that mass scale: the only way to grow more massive is by accretion of lower-mass galaxies. 

The mass growth of galaxies with $M_{\star}\approx10^{9.5}-10^{10.5}~\mathrm{M}_{\odot}$ is dominated by in-situ star formation. For these galaxies, mergers are typically gas-rich and a disc can re-form \citep[e.g.,][]{hammer05, hopkins09a}. Nevertheless, mergers play an important role in the assembly of the structure of these galaxies: those with a high spheroid-to-total fraction and a high concentration also have a significantly enhanced ex-situ stellar mass fraction (see Figure~\ref{fig:fexsitu_M_ST}). The stellar mass density profiles (Figure~\ref{fig:profiles}) reveal that the spheroidal mass fraction peaks in the centre, while the ex-situ fraction peaks in the outskirts. Furthermore, the ex-situ stellar mass budget is clearly dominated by minor mergers (Figure~\ref{fig:fexsitu_M}). 

All of these findings are consistent with the picture suggested by \citet{hernquist89}, where tidal effects during mergers may induce instabilities that can funnel a large amount of gas into the central region of a galaxy, thereby inducing a starburst that creates a spheroidal component. Additionally, based on zoom-in simulations, it has been suggested that misaligned accretion (i.e. counter-rotating gas accretion) can also lead to gas compaction and the formation of a central spheroidal component \citep{sales12, zolotov15,tacchella16_profile}. This channel is especially efficient when gas fractions are high, e.g., for local low-mass galaxies and galaxies at high redshifts. This shows that typical star-forming galaxies on the main sequence can form bulges in-situ, consistent with observations \citep{nelson16, nelson19_mm, tacchella15_sci, tacchella18_dust}.

\begin{figure}
	\includegraphics[width=\linewidth]{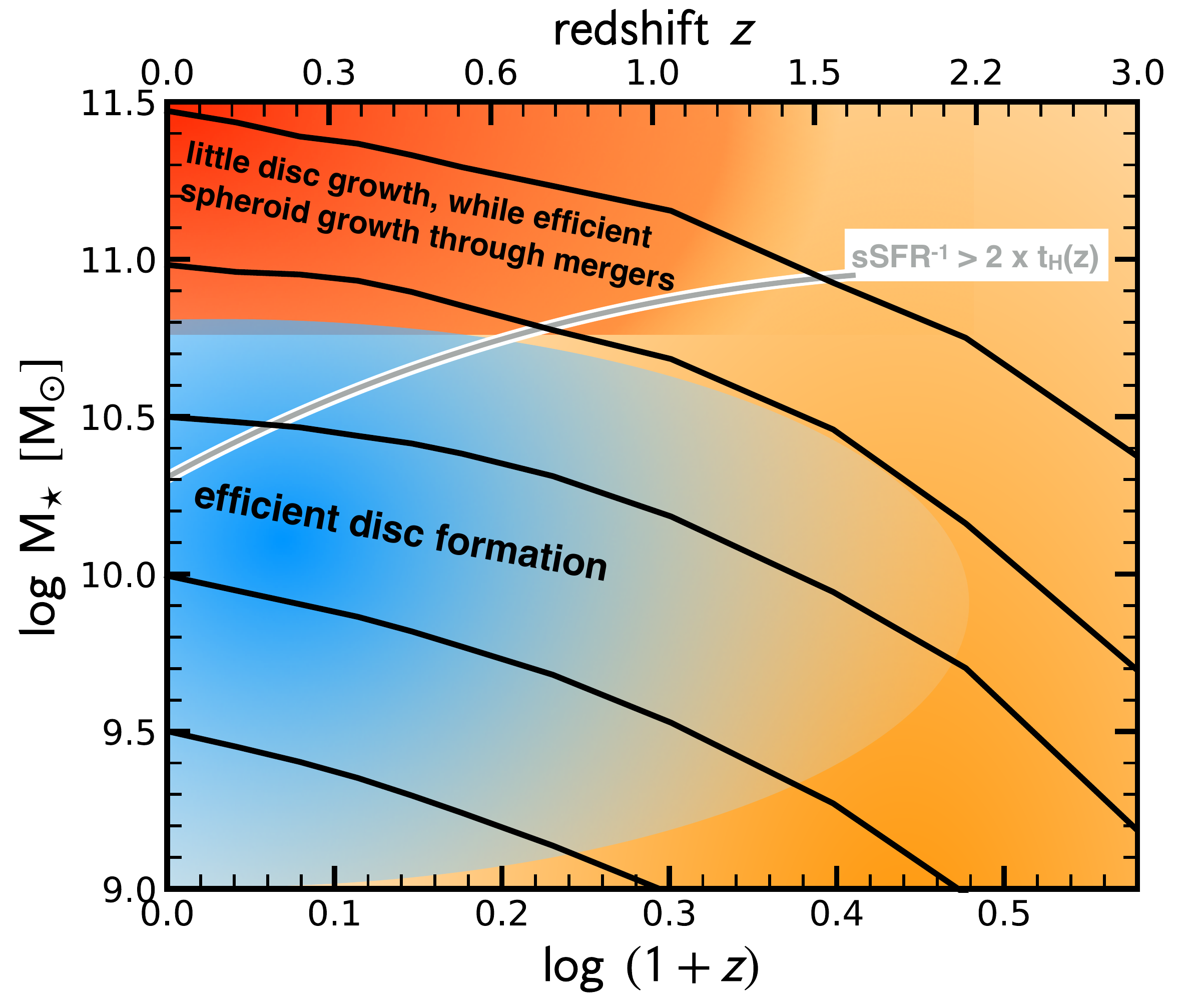}
    \caption{Sketch of the expected disc formation efficiency ($=\Delta M_{\rm disc}/\Delta M_{\star}$) in the plane of stellar mass ($M_{\star}$) and redshift. The solid black lines show the median mass growth histories of galaxies with $\log(M_{\star}(z=0)/\mathrm{M}_{\odot})=11.5$, 11.0, 10.5, 10.0, and 9.5. The green line indicates where the inverse of the sSFR, roughly the mass-doubling timescale, rises above twice the Hubble time. This line corresponds to the time when galaxies no longer efficiently increase $M_{\star}$ due to star formation. The efficiency for disc formation increases with cosmic time, in particular for stellar masses of $10^{9.5}-10^{10.5}~\mathrm{M}_{\odot}$. At lower masses, discs do not form efficiently, leading to high S/T values at low $M_{\star}$. Galaxies with $M_{\star}\ga10^{11}~\mathrm{M}_{\odot}$ at $z=0$ have never fully entered a phase of efficient disc formation. Their mass growth in recent times has been dominated by mergers that further build up the spheroidal component.}
    \label{fig:cartoon}
\end{figure}

The transition from the minor to major merger regime at $M_{\star}\approx10^{11}~\mathrm{M}_{\odot}$ is important. First, this transition is directly reflected in the stellar mass density profiles: the ex-situ mass fraction is radially increasing when minor mergers dominate, while it is roughly constant with radius when major mergers dominate (see Figure~\ref{fig:profiles}). Secondly, it can have important implications for the size growth of quiescent galaxies: minor mergers lead to the increase of the radius as the square of the mass in contrast to the usual linear rate of increase for major mergers \citep{naab09}. Finally, this mass scale is roughly consistent with the critical mass above which slow rotators are abundant \citep[e.g.,][]{cappellari16}. 

We summarize the emerging picture of spheroid and disc growth in IllustrisTNG in Figure~\ref{fig:cartoon}. Both $M_{\star}$ and cosmic time play crucial roles in understanding the build-up of morphology. Stellar mass and its correlation with star formation controls whether galaxies grow mainly by in-situ star formation or by ex-situ mergers. Cosmic time sets the gas accretion rate and gas content of the galaxies. As shown in \citet{pillepich19}, star-forming, H$\alpha$ emitting gas of TNG galaxies with stellar masses of $>10^9~\mathrm{M}_{\odot}$ at $z=0.5-4$ is always colder and in discier configurations than the stars. Hence, the stars that form initially cold are either quickly heated up into kinematically hot configurations, or the different star-formation events are misaligned, leading to an overall hot stellar system. 

In any case, significant spheroidal assembly takes place during the star-formation phase of the galaxies at early cosmic times. Galaxies are able to efficiently grow a disc at lower redshifts ($z<2$) and in the mass regime of $M_{\star}\approx10^{9.5}-10^{10.5}~\mathrm{M}_{\odot}$. The most massive galaxies have moved through this mass range too early and were thus unable to efficiently form a disc (Figure~\ref{fig:cartoon}). Their stellar component is at all times dominated by random motions. About half of their spheroidal component has assembled while they were star-forming at early times, the other half through mergers after ceasing their star formation nearly 6-10 billion years ago (Figure~\ref{fig:fraction_growth_Msph}). Those mergers directly build up the spheroid component, but also indirectly contribute by disrupting any pre-existing stellar disc (though this process is subdominant on average). 

From these considerations and from Figure~\ref{fig:evolutionary_tracks}, it is evident that galaxies do not necessarily need to change their morphology when they cease their star formation in order to explain the colour-morphology relation observed at $z=0$. Specifically, we show in Figure~\ref{fig:morphology_color} that, at fixed $M_{\star}$, S/T depends on the $(g-r)$ colour, while $C_{82}$ does not. This difference can be explained by the fact that the S/T$-M_{\star}$ evolves with cosmic time, while the $C_{82}-M_{\star}$ relation does not. Galaxies that quench and leave the star-forming main sequence at early times have, at that moment, a higher S/T value than similar-mass galaxies today, though their $C_{82}$ is comparable (see also Figure~\ref{fig:MS_morph}). The observed colour-S/T correlation is simply a reflection of the fact that galaxies that formed their stars earlier formed them in a more spheroidal-like manner. 

This conclusion does not imply that dry mergers are entirely subdominant in spheroid formation. As shown in Figures~\ref{fig:fraction_growth_Msph} and \ref{fig:evolutionary_tracks}, there is significant growth of the spheroid component after galaxies cease their star formation. On average, the mass in the spheroid component doubles and S/T slightly increases by up to 0.1.

\subsection{Low-mass galaxies in IllustrisTNG}
\label{subsec:lowM_galaxies}

We find in Figure~\ref{fig:morphology_vs_M} that TNG galaxies with $M_{\star}\approx10^9-10^{9.5}~\mathrm{M}_{\odot}$ have low concentration ($C_{82}\approx3$), i.e., that their stellar mass density is discy (S\'{e}risc index of $n\approx1$) while their stellar kinematics are dominated by random orbits ($\mathrm{S/T}\approx0.7$ at $z=0$, and higher at earlier cosmic times). 

Before interpreting these results, one has to worry that resolution effects could play a role in these galaxies. It is possible that S/T of these low-mass galaxies is overestimated in TNG because of the assumed equation of state \citep{springel03}, which might prevent the formation of thin discs at low masses \citep{benitez-llambay18}. We discuss the impact of resolution on our morphology indicators in Appenidx~\ref{app:resolution}. We conclude that resolution may have an effect on the exact numbers quoted, but that the overall trends with stellar mass are robust. Consistently, \citet{pillepich19} show that TNG50-2 (with a similar resolution of the TNG100 simulation analysed here) returns a qualitatively consistent picture with TNG50, but they also demonstrate that lower resolution generally underestimates the number of discy galaxies. Specifically, the growth of disc fractions of low-mass galaxies is underestimated by up to 20 percent while it is overestimated at the highest mass end. These findings are consistent with lower resolution imposing thicker disc heights. 

Additional support for our findings of high S/T for low-mass galaxies comes from observations, which underpin the upturn of S/T towards low-mass galaxies: \citet{zhu18} perform statistical modelling of stellar orbits of 300 local galaxies, confirming the increased prevalence of the hot, warm, and counter-rotating component towards low-mass galaxies (see Figure~\ref{fig:fraction_sph}). Along similar lines, \citet{wheeler17} show that $\sim80\%$ of the local dwarf galaxies have dispersion-supported rather than rotation-supported stellar motions. Finally, \citet{simons15} quote $M_{\star}\approx10^{9.5}~\mathrm{M}_{\odot}$ as the mass above which all star-forming galaxies form discs based on H$\alpha$ kinematics.

What is the physical cause for the rather large S/T values of low-mass galaxies? Works based on higher-resolution simulations, in particular zoom-in simulations, find similarly high S/T values. Using the FIRE-2 simulations, \citet{el-badry18} find that 15 out of their 17 galaxies with $M_{\star}<10^{9.5}~\mathrm{M}_{\odot}$ are dominated by random stellar motion and show no clear sign of the formation of a cold stellar disc. Only more massive galaxies with $M_{\star}>10^{10}~\mathrm{M}_{\odot}$ show a prominent disc component at $z=0$ \citep{garrison-kimmel18}. They argue that the reduced rotational support in their low-mass galaxies is due to stellar feedback driving non-circular motions in the gas, in combination with heating by the UV background which suppresses the accretion of high angular momentum gas. Similarly, works based on the VELA simulations \citep{ceverino14_radfeed} find that galaxies around $10^9~\mathrm{M}_{\odot}$ at $z>1$ have stellar mass and SFR distributions following roughly exponential disc profiles \citep{tacchella16_profile}, though these galaxies tend to be triaxial, prolate, and dispersion-dominated \citep{zolotov15, ceverino15b, tomassetti16}. These authors argue that dark matter dominates the gravitational budget, which leads to the reduced rotational support. Rotating discs in these simulations form only after gas compaction events, which lead to a high central stellar mass density \citep{tacchella16_profile, tomassetti16}. 

We argue that the formation mechanism of these dispersion-dominated, low-mass systems needs to be investigated in more detail. Since the star-formation histories in IllustrisTNG are less variable on short time scales than the ones in FIRE-2, stellar feedback driving non-circular motions in the gas is probably not their main formation mechanism in TNG. As shown in \citet{pillepich19}, star formation takes place in cold and discy configuration. Therefore, the initially cold stars are either quickly heated up into kinematically hot configurations, or the different star-formation events are misaligned with respect to each other, leading to an overall hot system \citep{sales12, danovich15}.

\subsection{Outstanding questions}
\label{subsec:outstanding}

As highlighted in the previous section, the resolution of our simulation is an important caveat. The TNG100 simulation used in this work is a state-of-the-art simulation that matches many current observational constraints successfully \citep[e.g.,][see \citealt{nelson19_dr} for a summary]{donnari19, diemer19}. We have shown that galaxy morphology, as measured by S/T and $C_{82}$, is also reasonably consistent with observed trends. Nevertheless, it is difficult to prove that the structure of galaxies in these simulations is converged. As discussed above and shown in Appendix~\ref{app:resolution}, S/T and $C_{82}$ change when decreasing the resolution. However, the main trends and dependencies on colour and stellar mass, and thus the key conclusions of this paper, are robust.

A related aspect of the simulation is the sub-grid models for star formation and feedback. The ISM model in IllustrisTNG pressurizes the gas and might therefore not be reliable in low-mass galaxies ($M_{\star}\la10^{9.5}~\mathrm{M}_{\odot}$). The agreement of our morphological measurements for those galaxies with observations and also with zoom-in simulations might arise for the wrong, unphysical reasons. However, \citet{pillepich19} tested the impact of the sub-grid pressure term and concluded that the vertical structure of star-forming discs negligibly depends on the parameter choices in the ISM model. 

In terms of the feedback implementation, the results for IllustrisTNG and the original Illustris simulations exhibit large differences (e.g.,  Figures~\ref{fig:app_resolution} and \ref{fig:app_profiles_orig}). Thus, quantities such as stellar mass profiles can constrain the feedback models, given a detailed comparison with observations. We compared the S/T and $C_{82}$ values as a function of $M_{\star}$ from TNG with observational data in Section~\ref{subsec:comparison_obs} and Figure~\ref{fig:fraction_sph}. We find overall good agreement and the main trends from the simulations are confirmed by observations. 

A more detailed comparison concerning the morphology of galaxies has been presented in \citet{rodriguez-gomez19} and \citet{huertas-company19}, where the simulations have been forward-modeled into the observational space. \citet{rodriguez-gomez19} find that the optical morphologies of IllustrisTNG galaxies are in good agreement with observations. In particular, the median trends with stellar mass of all the morphological, size, and shape parameters considered in that work lie within the $\sim1\sigma$ scatter of the observational trends, consistent with our findings (Figure~\ref{fig:fraction_sph}). This agreement is to be expected because our mass-based  concentration measurements are in qualitative agreement with the light-based ones from \citet{rodriguez-gomez19}, as shown in Appendix~\ref{app:diff_morph}. Furthermore, they show that IllustrisTNG does not produce a strong morphology-colour relation, i.e., that the red disc fraction is too high. This is consistent with the absence of a trend between $C_{82}$ and colour that we illustrate in Figure~\ref{fig:morphology_color}, though we also show there that there is a correlation between S/T and colour. As we have discussed above, one key reason for this is that S/T$-M_{\star}$ evolves with redshift, while $C_{82}-M_{\star}$ does not. We speculate that this points towards some missing physical mechanism that introduces a $z$-dependent $C_{82}-M_{\star}$ relation. For example, gas might not be efficiently transported to the central region in high-$z$ galaxies. Further investigation of the gas mass, stellar mass, and SFR density profiles are needed. Additionally, \citet{bustamante19} showed that by introducing a new sub-grid model for the black-hole spin evolution, they are able to alleviate some of the tension in the galaxy morphology-colour relation. The reason for this is that the kinetic feedback mode is activated predominantly in merging galaxies, in which major morphological changes take place as well. Understanding this link between the black hole sub-grid physics and morphology further is crucial and motivates future studies.   

Along similar lines, \citet{huertas-company19} study the morphology in TNG, finding that the majority of high-mass galaxies are discs rather than spheroids, in contrast to our findings. They inferred morphology from the light distribution and classified galaxies with a machine learning algorithm trained on visually classified SDSS galaxies, while our morphological measures are based on kinematics and the mass distribution. Beyond the complications of the machine learning approach, this highlights the importance of kinematical versus photometric morphological measures and comparisons.

While forward modelling the simulations into the observational space is key in order to perform an apples-to-apples comparison, adding another layer of free parameters on top of the galaxy formation model also increases modelling uncertainty. It is therefore important to properly marginalize over those parameters before concluding that the underlying galaxy formation model needs to be modified.



\section{Summary and Conclusions}
\label{sec:conclusion}

We have studied the relation between morphology, stellar mass, and star-formation activity in the IllustrisTNG galaxy formation model. We focus on TNG100 central galaxies with stellar masses in the range $10^{9-11.5}~\mathrm{M}_{\odot}$. We quantify galaxy morphology by a kinematical decomposition of the stellar component into a spheroid and a disc component (S/T) and by the concentration of the stellar mass density profile ($C_{82}$). 

Our conclusions are as follows.

\begin{itemize}
    \item Both S/T and $C_{82}$ exhibit a strong $M_{\star}$ dependence. Typical disc galaxies with a low concentration and cold stellar kinematics ($\mathrm{S/T}<0.5$) have a stellar mass of $M_{\star}\approx10^{10}-10^{10.5}~\mathrm{M}_{\odot}$. More massive galaxies are spheroids with hot kinematics and a high concentration, while low-mass galaxies have hot kinematics but a low concentration. These trends agrees well with observations (Figure~\ref{fig:fraction_sph}).
    \item In addition to stellar mass, the morphology of TNG galaxies correlates with their $(g-r)$ colour and star-formation activity (Figures~\ref{fig:morphology_col_mass} and \ref{fig:morphology_color}). Specifically, S/T increases from the blue cloud to the red sequence at fixed $M_{\star}$. Contrarily, $C_{82}$ shows no colour dependence and only depends on stellar mass. Since $C_{82}$ correlates with light-based morphological indicators (Appendix~\ref{app:diff_morph}), this is consistent with \citet{rodriguez-gomez19}, who found the morphology-colour correlation to be too weak in IllustrisTNG.
    \item A key result of this work is that galaxies do not necessarily change their morphology when they transition through the green valley on to the quiescent sequence (Figure~\ref{fig:evolutionary_tracks}), this depending on mass scale and morphological estimator. Galaxies' morphology (particularly $C_{82}$ and spheroidal and bulge mass fractions) is largely set before they cease their star formation. The observed correlation between S/T and colour (star formation) is mostly a reflection of the fact that galaxies that formed their stars earlier formed them mostly in a spheroidal-like manner: the S/T$-M_{\star}$ relation evolves with cosmic time to lower values (disc assembly). On the other hand, the $C_{82}-M_{\star}$ relation remains constant with cosmic time, leading no correlation between $C_{82}$ and colour. 
    \item The formation of disc and spheroidal components depends critically on stellar mass and cosmic time. At early times ($z\ga2$), the mass growth of all galaxies is dominated by in-situ star formation, usually forming significant spheroidal components. Towards lower redshift ($z\la2$), discs start to form efficiently in the regime of $M_{\star}\approx10^{9.5-10.5}~\mathrm{M}_{\odot}$ (Figure~\ref{fig:disk_growth}). 
    \item Above this mass scale, galaxies quench efficiently due to black-hole feedback, implying that further mass growth occurs via ex-situ accretion. Minor mergers contribute significantly to the accretion of mass (Figure~\ref{fig:fexsitu_M}) that is preferentially deposited in the outskirts (Figure~\ref{fig:profiles}), especially in galaxies below $M_{\star}\la10^{11}~\mathrm{M}_{\odot}$. For the most massive galaxies, major mergers contribute the majority of ex-situ stellar mass, which dominates at all radii including the central region (Figure~\ref{fig:profiles}). Dry mergers can also play a role in spheroid formation: during the quiescent evolution of these massive galaxies, the mass in the spheroid component doubles and S/T slightly increases by up to 0.1 (Figures~\ref{fig:fraction_growth_Msph} and \ref{fig:evolutionary_tracks}). 
    \item Intermediate-mass galaxies ($M_{\star} \approx 10^{10-11}~\mathrm{M}_{\odot}$) exhibit a correlation between the ex-situ stellar mass fraction with S/T and $C_{82}$ (Figure~\ref{fig:fexsitu_M_ST}). This trend is consistent with a picture where mergers induce an enhancement of centrally concentrated star-formation that lead to the formation of a spheroidal component and a high concentration. Therefore, the bulge component consists mostly of stars that formed in-situ (Figure~\ref{fig:profiles}), though its formation is, to a large degree, induced by mergers. In future work, it would be interesting to decompose the bulge formation contributed by true ``in-situ gas'' versus ``ex-situ gas'' brought in by mergers.
\end{itemize}

The main conclusion of our work is that although we find a correlation between star-formation activity and morphology at a given epoch, we do not find a direct link between the two. In IllustrisTNG, the main mechanism responsible for the cessation of star formation is black-hole feedback, which itself has no direct dependence on morphology (neither on the kinematics of the stars nor on the concentration of the stellar mass profile). Clearly, both the morphology and the black-hole mass are shaped by similar physical processes such as mergers that can funnel gas into the central region. Nevertheless, the main reason for the morphology-star formation correlation is that the quiescent population contains galaxies that have ceased their star formation over a wide range of epochs, during which the star-forming galaxy population has evolved tremendously. Specifically, many of today's quiescent galaxies stopped their star formation when star-forming galaxies had a larger spheroidal mass fraction.

We have shown how a numerical simulation such as TNG can be used to help to interpret observations and guide future observing endeavours. As highlighted throughout the text, we have left several follow-up studies for future work, including a more detailed comparison with observations. In particular, we will learn more about the low-mass regime from the TNG50 simulation \citep{nelson19, pillepich19}, which has a mass resolution that is about 16 times better than the simulation analysed here. 

\section*{Acknowledgements}

We thank the referee for carefully reading the manuscript and providing constructive feedback. We are grateful to Mauro Giavalisco, Susan Kassin, Rohan Naidu, Erica Nelson, and Paul Torrey for productive discussions. This research made use of NASA's Astrophysics Data System (ADS), the arXiv.org preprint server, and the python packages \texttt{Matplotlib} \citep{hunter07}, \texttt{Astropy} \citep{astropycollaboration13}, and \texttt{Colossus} \citep{diemer18_colossus}. The TNG100 simulations used in this work have been run on the HazelHen Cray XC40-system at the High Performance Computing Center Stuttgart as part of project GCS-ILLU of the Gauss Centres for Supercomputing (GCS). ST is supported by the Smithsonian Astrophysical Observatory through the CfA Fellowship. BD acknowledges support from the program HST-HF2-51406.001-A, which was provided by NASA through a grant from the Space Telescope Science Institute, operated by the Association of Universities for Research in Astronomy, Incorporated, under NASA contract NAS5-26555. FM acknowledges support through the Program ``Rita Levi Montalcini'' of the Italian MIUR. VRG acknowledges support from the National Science Foundation under Grant No. AST-1517559. MV acknowledges support through an MIT RSC award, a Kavli Research Investment Fund, NASA ATP grant NNX17AG29G, and NSF grants AST-1814053 and AST-1814259.







\appendix

\section{Resolution dependence in TNG}
\label{app:resolution}

\begin{figure}
	\includegraphics[width=\columnwidth]{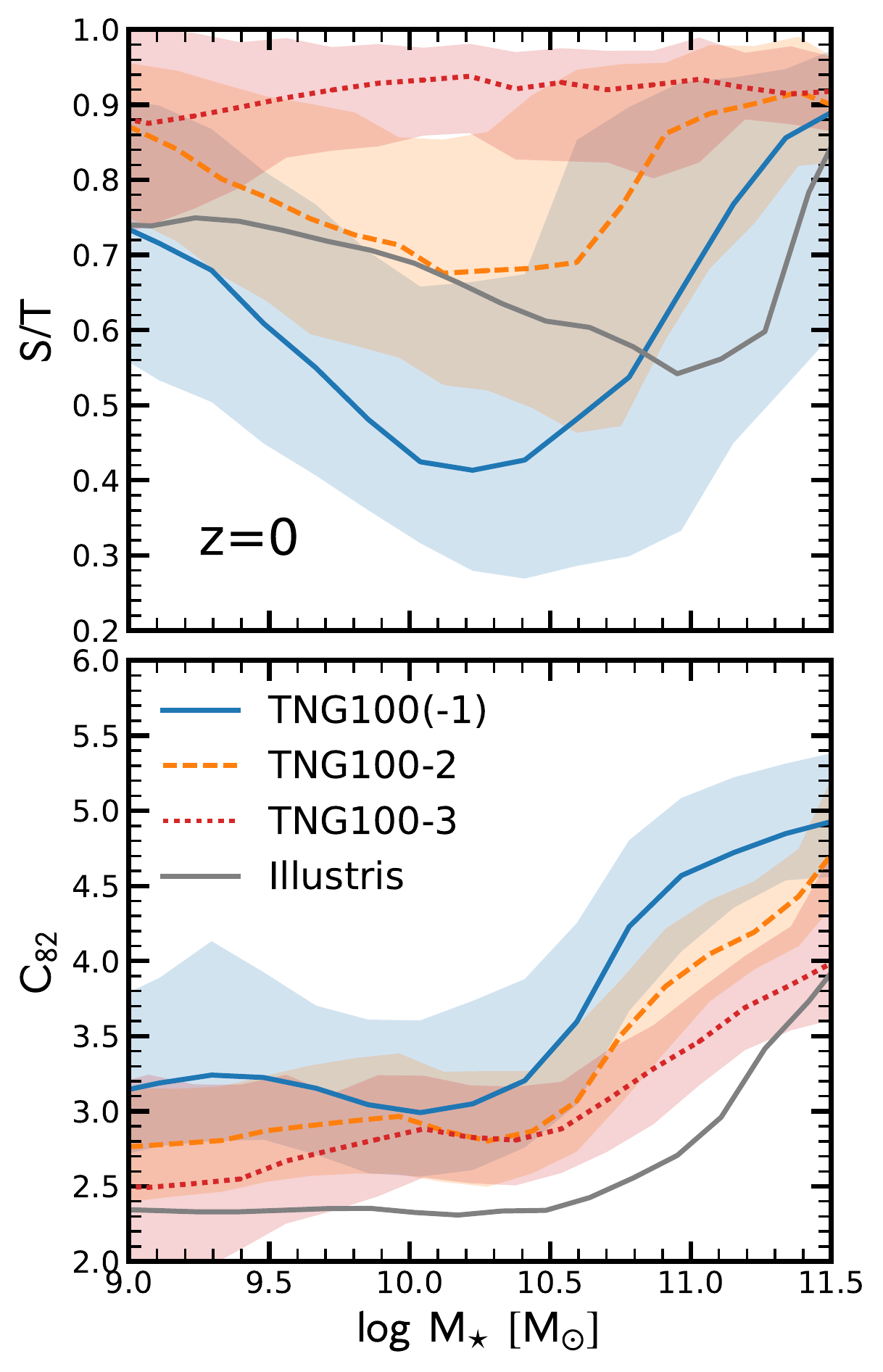}
    \caption{Effects of numerical resolution on morphological indicators S/T (upper panel) and $C_{82}$ (lower panel). In both panels we plot the median relations and 16th/84th percentiles as a function of stellar mass. The solid blue, dashed orange, and dotted red lines indicate the TNG100(-1), TNG100-2, and TNG100-3 simulations, respectively (see Table~\ref{tab:TNG100}). As a comparison, we show as gray solid line the original Illustris simulation. }
    \label{fig:app_resolution}
\end{figure}

We consider here effects of resolution in the IllustrisTNG simulations. Throughout this paper we have exclusively shown results from our fiducial TNG100-1 volume, whereas here we consider the effects of changing resolution. To do so we analyse TNG100-2 and TNG100-3, realizations with 8 and 64 times lower mass resolution, corresponding to factors of 2 and 4 larger gravitational softening lengths. The numerical parameters of this resolution series are given in Table~\ref{tab:TNG100}. 

\citet{pillepich18} discuss the effects of resolution and numerical convergence in detail. Briefly, using this resolution series, they find higher resolution results in larger galaxy stellar masses at fixed halo mass, leading to a higher normalization of the stellar mass function and cosmic SFR density. Many of the other galaxy population statistics exhibit a similar qualitative trend, i.e. monotonically larger values with increasing numerical resolution. This can be understood by reference to the implementation of star formation in the TNG model. Progressively better spatial resolution leads to the sampling of ever higher gas density regions, allowing more gas mass to become eligible for star formation and to be resolved at higher densities, accelerating the rate at which it is turned into stars.

Using the resolution series of TNG100 (Table~\ref{tab:TNG100}), in Figure~\ref{fig:app_resolution} we explore the convergence properties of morphologies in the TNG model. Specifically, we plot the median S/T and $C_{82}$ versus $M_{\star}$ relations. The solid blue line shows our fiducial model, while the dashed orange and dotted red lines indicate the lower resolution versions. Furthermore, the grey solid line is the original Illustris model. 

Reducing the resolution leads to the disappearance of kinematically defined disc galaxies. Since discs are most prominent at intermediate masses ($M_{\star}\approx10^{9.5}-10^{10.5}~\mathrm{M_{\odot}}$), the median S/T changes most significantly for those galaxies. However, when analysing the scatter in Figure~\ref{fig:app_resolution}, we see that also discs disappear at lower and higher masses when decreasing the resolution. Consistent with this, \citet{pillepich19} show that TNG50-2 (comparable resolution of TNG100 analysed here) return a qualitatively consistent picture with TNG50 but demonstrate that lower resolution generally underestimates the number of disc galaxies. Specifically, the growth in time of the disc fractions of low-mass galaxies is underestimated (by up to 20 percent), while it is overestimated at the highest mass end. This is consistent with lower resolution imposing thicker disc heights.

The concentration seems to be less affected by resolution. The largest effect concerns high-mass galaxies, where decreasing the resolution leads to more diffuse galaxies. At lower masses, a similar trend can be seen, but the overall effect is less severe than at higher masses. We have also check the actual values for $r_{20}$, the radius that encloses $20\%$ of the stellar mass, and check it they are comparable to the gravitational softening length of the simulation. We find that only $16\%$ of the galaxies below $10^{9.5}~M_{\odot}$ have $r_{20}$ to be smaller than the gravitational softening. We conclude that this is not the cause of the flattening seen in the $C_{82}-M_{\star}$ relation at low stellar masses.

In Figure~\ref{fig:app_resolution}, we also compare the TNG results with the original Illustris simulations, where we have measured all the properties in exactly the same way. Illustris has practically the same resolution as the TNG100-1 simulation (our fiducial run). Nevertheless, we can find significant differences between these simulations. First, the original Illustris simulation produces much fewer disc-dominated galaxies than TNG, in particular in the mass regime of $10^{9.5}-10^{10.5}~\mathrm{M}_{\odot}$. This most probably arises because of the improved stellar feedback in TNG. At high masses ($>10^{11}~\mathrm{M}_{\odot}$), the upturn in S/T is taking place at lower masses in TNG than in the original Illustris simulation, which can be explained by the fact that more galaxies were star forming at those mass scales in Illustris. Through the improved black-hole feedback, galaxies are quenching more effectively at this mass scale in TNG. Similarly, the concentrations are lower in original Illustris and the turnover towards higher concentrations happens at higher stellar masses than in TNG. 

We conclude that, using the resolution series of TNG100, these simulations are not converged in an absolute sense, i.e. the physical properties of the galaxies change when increasing or decreasing the resolution. These changes are significant, though the qualitative trends remain. We are therefore confident that we grasp the main evolutionary trends of the galaxy population concerning their morphology and star formation activity.

\section{Comparing different morphological indicators}
\label{app:diff_morph}

We consider here how different definitions of morphology are correlated with each other. Specifically, in the main text we focus on two morphological indicators: S/T and $C_{82}$. In this section, we compare how these indicators change when adopting a different definition of the spheroidal component and when one performs the measurements on the light instead of the mass distribution.

\subsection{Different spheroid definitions}

\begin{figure*}
	\includegraphics[width=\textwidth]{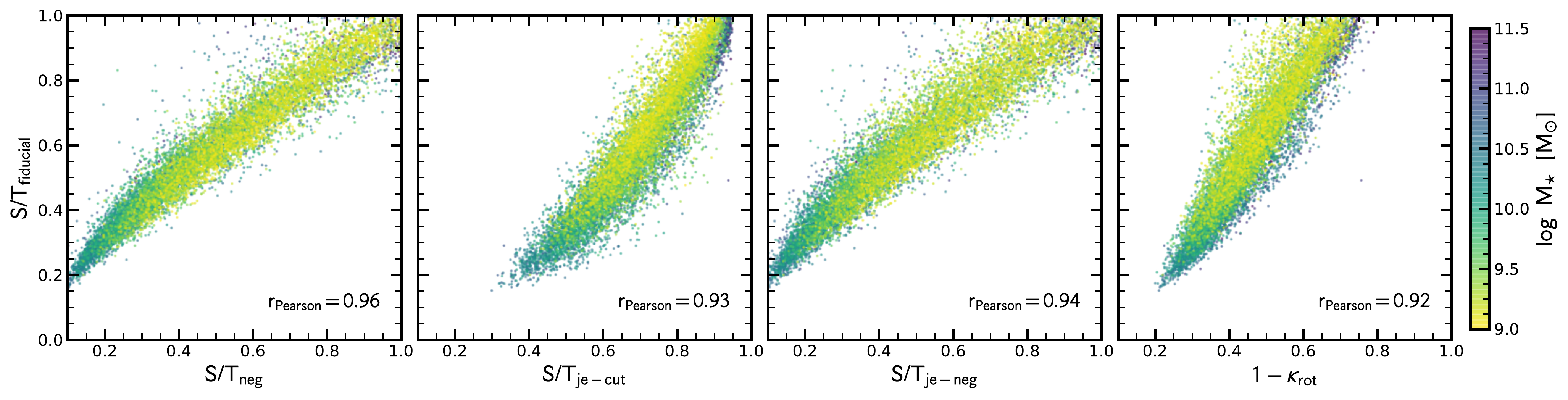}
    \caption{Good agreement between different spheroid definitions. Our fiducial S/T, based on $j_{\rm z}/j<0.7$, is compared to (i) the one defined by twice the mass of all stellar population particles that counter-rotate (left panel); (ii) by defining the rotation as $j_{\rm z}/j(E)$ where $j(E)$ is the angular momentum of a particle on a circular orbit with the same binding energy (middle left and middle right panels), (iii) $\kappa_{\rm rot}$, which is the fraction of the total kinetic energy contributed by the azimuthal component of the stellar angular momenta (right panel). All of these definitions lead to similar S/T values that are highly correlated with Pearson correlation coefficients $>0.9$.}
    \label{fig:app_diff_sph}
\end{figure*}

We have tested a number of kinematic spheroid-disc decompositions as well as the fraction of kinetic energy in rotation, $\kappa_{\rm rot}$, as defined in \citet{sales12} and \citet{rodriguez-gomez17}. In our fiducial calculation, we compute the spheroidal component as the mass of all stellar particles with $j_{\rm z}/j<0.7$ plus a 15\% correction as discussed in the main text. Here, $j$ is the total angular momentum of the particle and $j_{\rm z}$ the part that is aligned with the galaxy's rotation (as defined by the total angular momentum of all stellar particles bound to the galaxy). We have also experimented with alternative ways to quantify the spheroidal component: (i) S/T$_{\rm neg}$: twice the mass of all stellar population particles that counter-rotate, where rotation is defined as $j_{\rm z}/j$; (ii) S/T$_{\rm je-cut}$: the mass of all stellar particles with $j_{\rm z}/j(E)<0.7$, where $j(E)$ is the angular momentum of a particle on a circular orbit with the same binding energy \citep[e.g.,][]{marinacci14}; and (iii) S/T$_{\rm je-neg}$: twice the mass of all stellar population particles that counter-rotate, where rotation is defined as $j_{\rm z}/j(E)$.

Another kinematic morphology tracer is $\kappa_{\rm rot}$ \citep{sales12}, defined as the fraction of the total kinetic energy contributed by the azimuthal component of the stellar angular momenta, where the $z$-axis coincides with the total angular momentum of the galactic stellar component:

\begin{equation}
    \kappa_{\rm rot} = \frac{K_{\rm rot}}{K} = \frac{1}{K} \sum_{i} \frac{1}{2} m_i \left( \frac{j_{z,i}}{R_i} \right)^2
\end{equation}

\noindent
where $K$ is the total kinetic energy of the stellar component, $m_i$ represents the mass of a particle, $R_i$ is the projected radius and the sum is carried out over all stellar particles in the galaxy (i.e. not just within $3\times R_{\rm M}$). Following \citet{genel15}, the calculation frame is centred at the position of the most bound stellar particle, while the velocity of the frame coincides with that of the stellar centre of mass.

We compare these different definitions in Figure~\ref{fig:app_diff_sph}. All the different definitions for S/T correlate very strongly with each other (Pearson correlation coefficient of $>0.9$), indicating that the exact method for calculating S/T is not important for our purposes. In addition to scatter, we find small offsets on average between the different S/T estimates. Specifically, the S/T$_{\rm neg}$ estimates are on average lower than the fiducial S/T estimates. However, the overall change is small and nearly mass independent, leading to a simple overall shift when studying the S/T$-M_{\star}$ relation.

\subsection{Kinematics versus mass versus light}

\begin{figure}
	\includegraphics[width=\columnwidth]{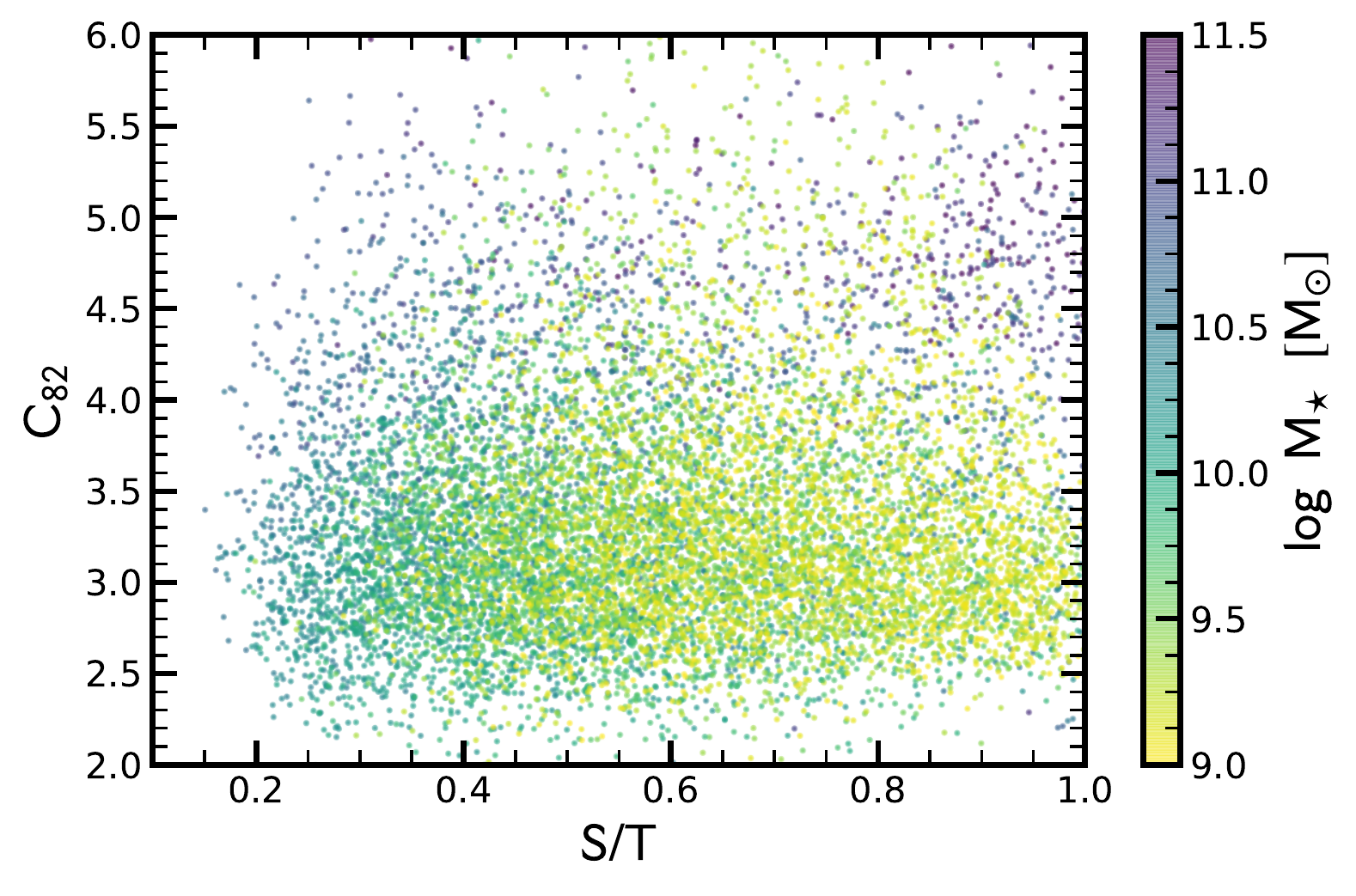}
    \caption{Stellar mass density versus kinematic morphology. We plot the concentration versus the spheroid-to-total ratio, colour-coded by the stellar mass. We find a large amount of scatter, indicating that these morphological indicators trace different structure within galaxies.}
    \label{fig:app_C_vs_ST}
\end{figure}

\begin{figure*}
	\includegraphics[width=\textwidth]{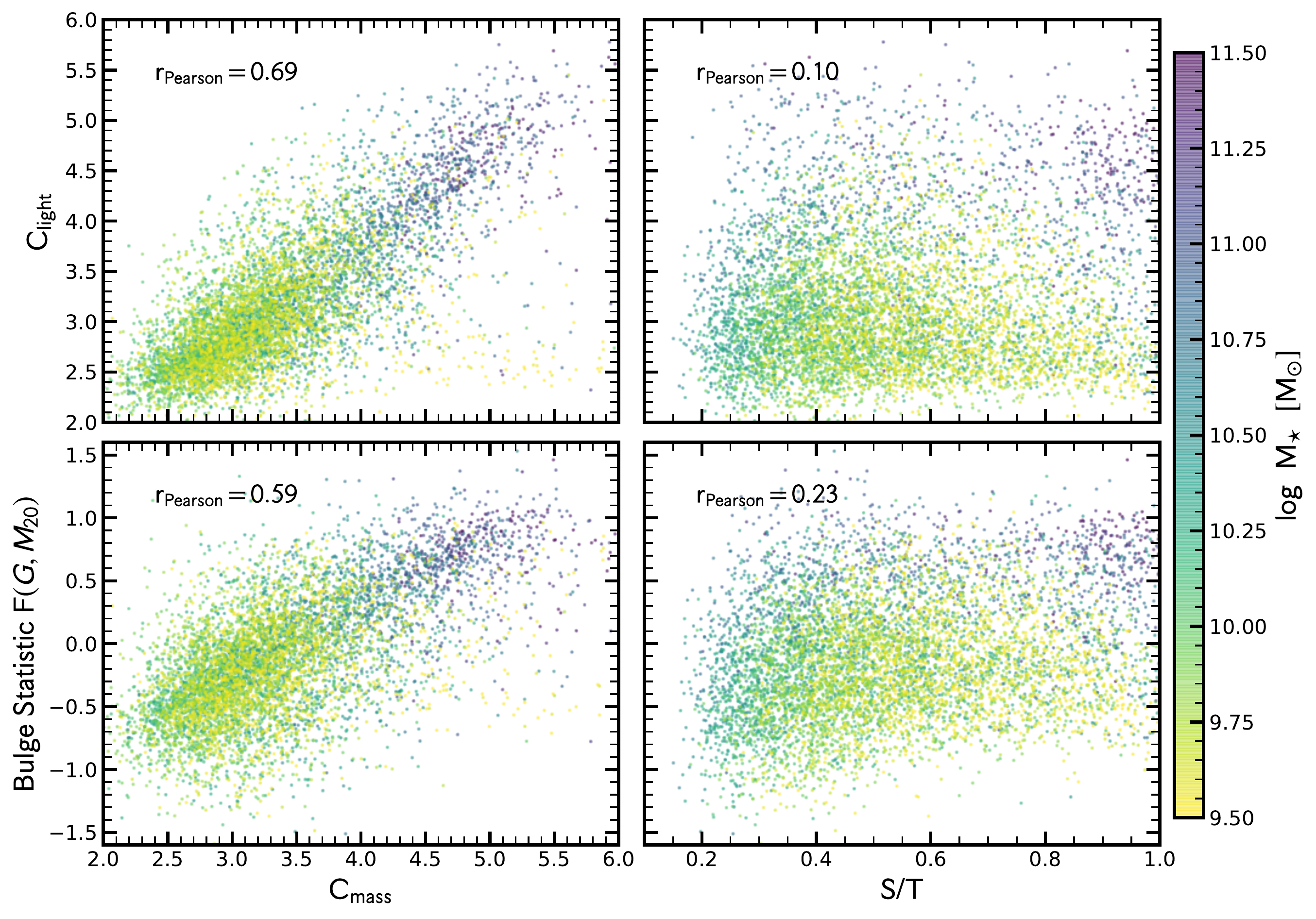}
    \caption{Relation between mass-based, light-based and kinematic measurements of morphology. The concentration ($C_{\rm light}$) and the bulge statistics ($F(G, M_{20})$) based on the light distribution in the galaxies are compared to the concentration of the stellar mass profile ($C_{\rm mass}$) and the kinematic S/T. We find that the mass-based concentration is correlated with the light-based concentration and bulge statistic, while the kinematic S/T is not.}
    \label{fig:app_mass_vs_light}
\end{figure*}

As shown in Figure~\ref{fig:morphology_col_mass}, S/T and $C_{82}$ depend differently on stellar mass and star formation. Here, we plot in Figure~\ref{fig:app_C_vs_ST} the relation between S/T and $C_{82}$. As expected, there is a large amount of scatter and no correlation visible. The major part of the scatter can be attributed to galaxies with low stellar masses ($M_{\star}<10^{10}~\mathrm{M}_{\odot}$). Those galaxies have typically a low concentration, i.e. are disc-like concerning their stellar mass distribution, but have a significant spheroidal component concerning their kinematics. More massive galaxies ($M_{\star}>10^{10}~\mathrm{M}_{\odot}$) reveal a tighter correlation with a Pearson correlation coefficient of 0.54. Overall, Figure~\ref{fig:app_C_vs_ST} highlights that the morphology according to the mass distribution can be quite different from the kinematic morphology. 

The main focus of this paper is the investigation of the assembly and evolution of spheroidal and disc components of TNG galaxies. Hence, we work mainly in theory plane, i.e. quantify the morphology of galaxies based on the stellar mass distribution and kinematics. We only perform a rough comparison to observations in Section~\ref{subsec:comparison_obs}. One way to achieve a more detailed comparison to observations is based on ``forward-modelling'' of simulation data into the observational plane. In particular, the generation and subsequent analysis of synthetic images from hydrodynamic simulations is a powerful tool to connect theory with observations \citep[e.g.,][]{jonsson10, scannapieco10, snyder15,snyder15_morph,bottrell17}. However, the generation of realistic synthetic images is challenging such as the aspects of the detailed modelling of dust absorption and scattering.

In \citet{rodriguez-gomez19}, we created realistic synthetic images from the IllustrisTNG simulations using the radiative transfer code \texttt{SKIRT} \citep{baes11,camps15}. Specifically, we designed them to match  observations of low-redshift galaxies from the Pan-STARRS $3\pi$ Survey \citep{chambers16}. We then quantified various structural parameters of both the simulated and observed galaxies using the same morphology code, which allows us to make a fair comparison. We find that the optical morphologies of IllustrisTNG galaxies are in good agreement with Pan-STARRS observations: the locus of the Gini$-M_{20}$ diagram is consistent with that inferred from observations, while the median trends with stellar mass of all the morphological, size and shape parameters lie within the 1$\sigma$ scatter of the observational trends. However, the TNG model has some difficulty with more stringent tests, such as producing a strong morphology-colour relation. This results in a somewhat higher fraction of red discs and blue spheroids compared to observations.

We compare the mass-based and kinematic morphology indicators to light-based indicators in Figure~\ref{fig:app_mass_vs_light} in order to assess how similar they are. In particular, we compare our mass-based concentration and S/T estimates with the light-based concentration and the ``bulge statistic'' for individual galaxies at $z=0.05$. The bulge statistic is based on a combination of the Gini and $M_{20}$ estimates and quantifies the strength of the bulge component, and hence is tightly correlated with the light-based concentration \citep{snyder15_morph}. 

We find that the mass-based concentration is correlated with both the light-based concentration and the bulge statistic (Figure~\ref{fig:app_mass_vs_light}). The correlation is quite strong with a Pearson correlation coefficient of $0.6-0.7$. We speculate that the scatter is primarily caused by the effects of (spatially resolved) dust attenuation and scattering, which typically reduces the concentration, though also radial variation in the mass-to-light and spherical versus elliptical apertures could play a role. On the other hand, there is no correlation between these light-based indicators and the kinematic morphological tracer S/T. We find that the light-based concentration and bulge statistic are systematically lower than the kinematic fractions at low masses, but with increasingly good agreement as the stellar mass increases. This is similar to the findings presented in \citet{scannapieco10} and  \citet{bottrell17b}. We conclude that the absence of the strong morphology-colour relation seen in \citet{rodriguez-gomez19} is a direct reflection of our finding that there is no $C_{82}$-colour relation in Figure~\ref{fig:morphology_color}. 

\section{Morphology on the star-forming main sequence}
\label{app:MS_morph}

\begin{figure*}
	\includegraphics[width=\textwidth]{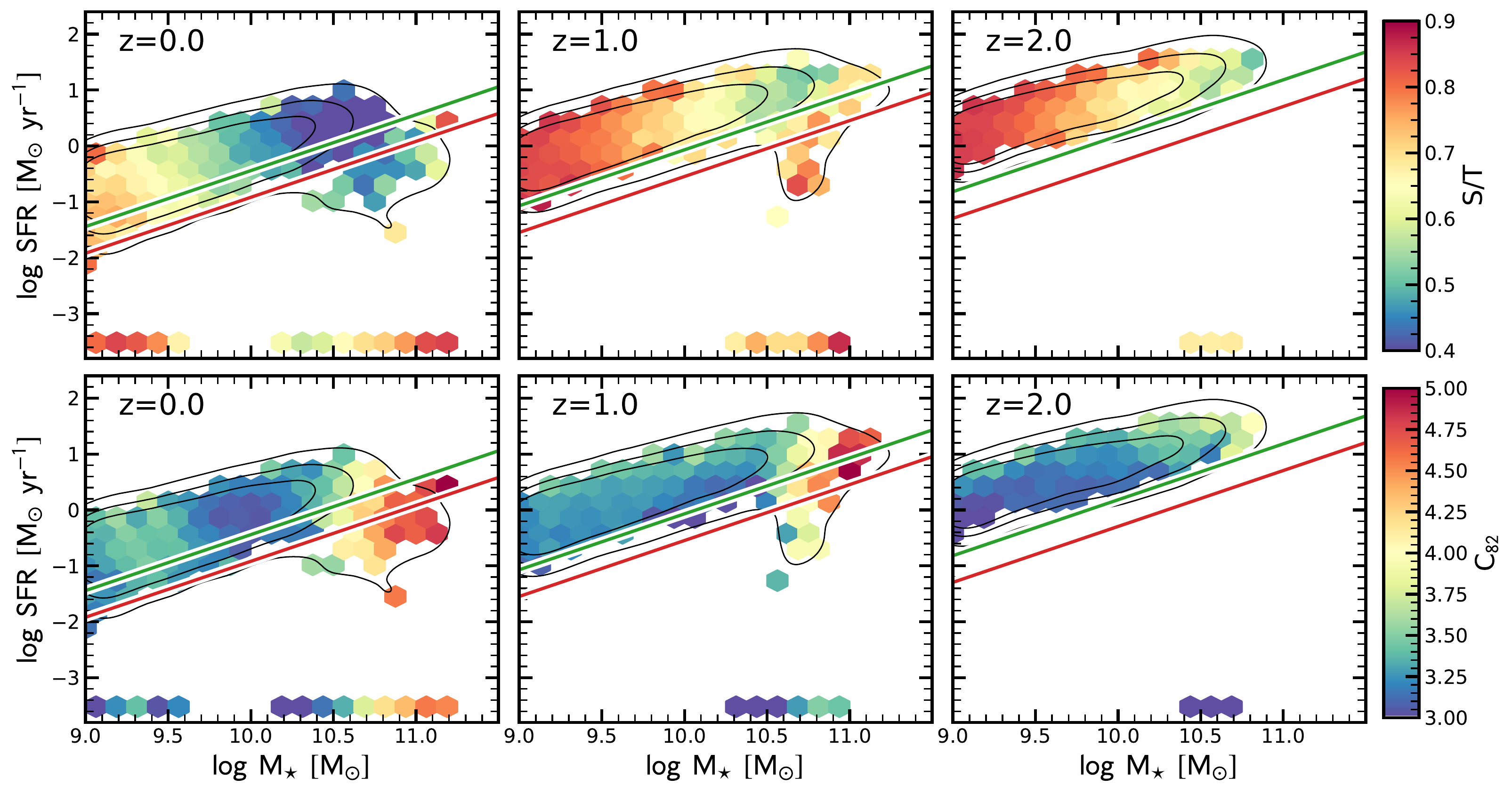}
    \caption{Relation between morphology, stellar mass and star-formation activity. We plot S/T (top) and $C_{82}$ in the plane of SFR and $M_{\star}$. The layout of this figure follows closely Figure~\ref{fig:morphology_col_mass}: each bin is coloured according to the median S/T or $C_{82}$ value as indicated by the colour bars on the right and encloses at least 10 galaxies. Galaxies with $\mathrm{SFR}<10^{-3.5}~\mathrm{M_{\odot}}/\mathrm{yr}$ are set to $\mathrm{SFR}=10^{-3.5}~\mathrm{M_{\odot}}/\mathrm{yr}$. The left, middle and right panels show the galaxy population at $z=0$, $z=1$, and $z=2$, respectively. The green and red lines divide the galaxies into star-forming and transitional, and transitional and quiescent objects according to Equations~\ref{eq:ssfr_quiescent} and \ref{eq:ssfr_transition}.}
    \label{fig:MS_morph}
\end{figure*}

Figure~\ref{fig:MS_morph} shows the distribution of morphology (S/T and $C_{82}$) in the plane of SFR and $M_{\star}$, i.e. on the star-forming main sequence. Since SFR and $(g-r)$ colour are closely related, this figure is consistent with the conclusions drawn from Figure~\ref{fig:morphology_col_mass}. The green and red lines mark the transition and quiescent regions according to the Hubble time as detailed in Equations~\ref{eq:ssfr_quiescent} and \ref{eq:ssfr_transition}. Since the normalization of star-forming main sequence roughly evolves proportional to the Hubble time, these cuts are similar to cuts relative to the star-forming main sequence. 

As discussed in the main text, there are significant differences between S/T and $C_{82}$. First, while $C_{82}$ mainly depends on $M_{\star}$, S/T shows an $M_{\star}$ and SFR dependence. Secondly, the $M_{\star}$ dependence of S/T and $C_{82}$ are different, in particular for low-mass galaxies. Finally, there is a much stronger trend with redshift for S/T than for $C_{82}$. In particular, we see that disc galaxies with low S/T only appear in significant numbers at $z<1$. This is important since it directly translates into a correlation between S/T and colour, while there is no correlation between $C_{82}$ and colour: quiescent galaxies today were star-forming in the past when S/T were overall higher.

\section{Stellar mass profiles in original Illustris}
\label{app:illustris}

\begin{figure*}
	\includegraphics[width=\textwidth]{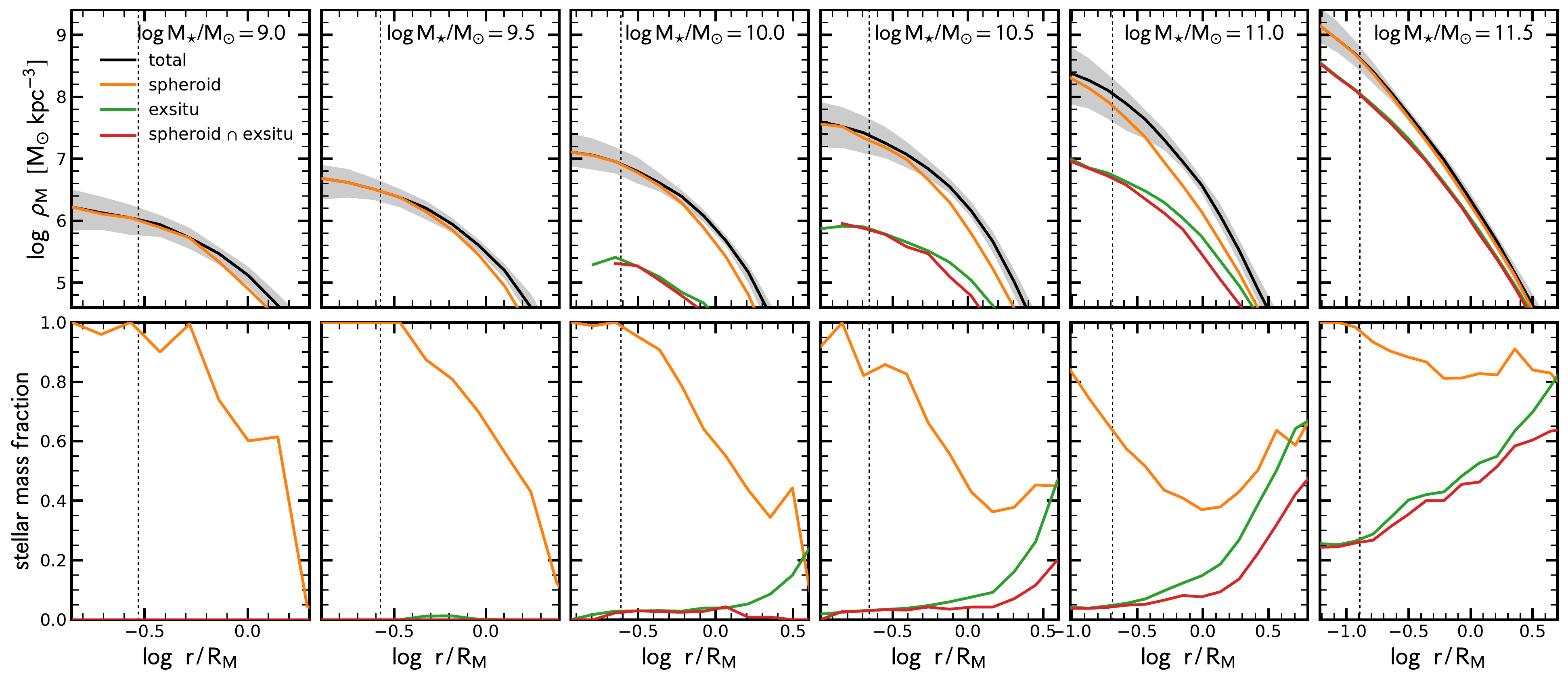}
    \caption{Stellar mass density profiles measured in the original Illustris simulation. This figure follows exactly the same layout as Figure~\ref{fig:profiles}. The panels from the left to the right show increasing stellar mass bins from $M_{\star}\approx10^{9}$ to $10^{11.5}~\mathrm{M}_{\odot}$. The top panels show the 3D stellar mass density profile: the black, orange, green and red lines indicate the profiles for the total stellar mass, the spheroidal component, the ex-situ component, and the spheroid\&ex-situ component, respectively. The radial coordinate of all profiles are normalized by the stellar half-mass radius. The bottom panels show the fraction contribution of the spheroidal, ex-situ and spheroid\&ex-situ component to the total stellar mass density profile. }
    \label{fig:app_profiles_orig}
\end{figure*}

Figure~\ref{fig:app_profiles_orig} plots the stellar mass density profiles as measured in the original Illustris simulation. The figure follows the exact same layout as Figure~\ref{fig:profiles} in the main text, where we show the profiles for the TNG simulation. Comparing the new TNG profiles with the ones from the original Illustris reveals several differences. First and foremost, the profile shapes are very different: the galaxies in TNG are significantly more concentrated at all stellar masses. This is consistent with the findings of Figure~\ref{fig:app_resolution}. Secondly, the spheroidal component is more dominant in the central region in the original Illustris-1 simulation than in the TNG simulation, in particular in intermediate galaxies ($M_{\star}\approx10^{10}-10^{10.5}~\mathrm{M}_{\odot}$). Finally, the ex-situ fraction for massive galaxies is showing a steep gradient with radius in Illustris-1, while the this trend is much weaker in TNG: at $M_{\star}\approx10^{11.5}~\mathrm{M}_{\odot}$, the ex-situ stellar mass fraction dominates only in the outskirts (outside of $\sim3~R_{\rm M}$) in Illustris-1, while it dominates at all radii in TNG. 

These trends can be explained by the fact that the galaxies in Illustris-1 are more diffuse and have larger sizes than the galaxies in TNG. By construction, the galaxy sizes in TNG are in much better agreement with observations than the ones in the original Illustris-1 \citep{genel18,pillepich18}. Since more diffuse galaxies disperse more efficiently in a galaxy-galaxy merger, mergers in Illustris-1 are not able to bring significant amounts of stellar mass to the centre, which leads to a high ex-situ stellar mass fraction in the outskirts, but a low fraction in the centre. This highlights that there is a direct link between the morphology of galaxies and merger products.


\bsp	
\label{lastpage}
\end{document}